\pdfoutput=1
\documentclass{article}

\usepackage{PRIMEarxiv}

\usepackage[utf8]{inputenc}
\usepackage[T1]{fontenc}
\usepackage{amsmath}
\usepackage{booktabs}
\usepackage{array}
\usepackage{microtype}
\usepackage{graphicx}
\usepackage{xcolor}
\usepackage{fancyhdr}
\usepackage{tikz}
\usetikzlibrary{arrows.meta,positioning,fit,calc}
\usepackage[round,authoryear]{natbib}
\usepackage{url}
\usepackage[hidelinks]{hyperref}
\graphicspath{{media/}}
\newcommand{\NAttacks}{563,349}
\newcommand{\NNonSelf}{415,644}
\newcommand{\NAccounts}{5,329}
\newcommand{\NAttackers}{3,244}
\newcommand{\NTargets}{4,415}
\newcommand{\NDefenses}{118,377}
\newcommand{\NDeployed}{53,974}
\newcommand{\IdenticalShare}{11}
\newcommand{\DefenseStart}{July 24, 2023}
\newcommand{\NTexts}{271,378}
\newcommand{\NFamilies}{121,226}
\newcommand{\LargestFamily}{1,667}
\newcommand{\SuccessRateNonSelf}{7.9}
\newcommand{\ShareGPT}{78}
\newcommand{\ShareClaude}{13}
\newcommand{\SharePaLM}{8}
\newcommand{\FirstDate}{June 17, 2023}
\newcommand{\LastDate}{February 17, 2024}
\newcommand{\NEventsAll}{251,999}
\newcommand{\NEventsCtrl}{250,465}
\newcommand{\NEvents}{245,829}
\newcommand{\NEventVictims}{3,910}
\newcommand{\NControls}{1,220,305}
\newcommand{\MeanControls}{5.0}
\newcommand{\NExpSucc}{13,839}
\newcommand{\RateSucc}{10.0}
\newcommand{\NExpLeak}{7,164}
\newcommand{\RateLeak}{7.4}
\newcommand{\NExpNear}{9,621}
\newcommand{\RateNear}{1.0}
\newcommand{\NExpFail}{215,205}
\newcommand{\RateFail}{0.72}
\newcommand{\RateCtrl}{0.03}
\newcommand{\RateRecv}{1.45}
\newcommand{\RatioRecv}{45}
\newcommand{\WithinTextFail}{0.75}
\newcommand{\WithinTextCtrl}{0.034}
\newcommand{\ShareWithinDay}{81}
\newcommand{\HOneA}{1.42}
\newcommand{\HOneB}{2.06}
\newcommand{\HOneC}{1.42}
\newcommand{\HOneLeakC}{7.6}
\newcommand{\HOneD}{0.46}
\newcommand{\HOneFailD}{\ensuremath{-}0.74}
\newcommand{\HOneLeakD}{6.5}
\newcommand{\HOneDlo}{0.25}
\newcommand{\HOneDhi}{0.68}
\newcommand{\SuccMin}{8.8}
\newcommand{\SuccMax}{10.0}
\newcommand{\PresentInterTB}{5.6}
\newcommand{\PresentInterTSE}{0.4}
\newcommand{\AbsentRecvTB}{\ensuremath{-}0.2}
\newcommand{\AbsentRecvTSE}{0.1}
\newcommand{\PresentShare}{17}
\newcommand{\PresentSuccB}{37.2}
\newcommand{\PresentSuccSE}{2.7}
\newcommand{\AbsentSuccB}{4.6}
\newcommand{\AbsentSuccSE}{0.5}
\newcommand{\HTwoSuccB}{7.3}
\newcommand{\HTwoSuccLo}{6.2}
\newcommand{\HTwoSuccHi}{8.5}
\newcommand{\HTwoLeakB}{6.5}
\newcommand{\HTwoNearB}{\ensuremath{-}0.5}
\newcommand{\HTwoNearLo}{\ensuremath{-}1.6}
\newcommand{\HTwoNearHi}{0.7}
\newcommand{\SuccNearB}{7.8}
\newcommand{\SuccNearLo}{6.1}
\newcommand{\SuccNearHi}{9.6}
\newcommand{\HTwoTextSuccB}{9.8}
\newcommand{\NearStrictB}{\ensuremath{-}0.7}
\newcommand{\NearStrictLo}{\ensuremath{-}2.1}
\newcommand{\NearStrictHi}{0.6}
\newcommand{\NearStrictShare}{72}
\newcommand{\HTwoSelfSuccB}{7.3}
\newcommand{\FamNewSuccB}{2.6}
\newcommand{\FamNewNearB}{2.2}
\newcommand{\FamNewNearSE}{0.6}
\newcommand{\FamNewPriorRecv}{47}
\newcommand{\FailMeanEst}{2.0}
\newcommand{\FailMeanAll}{0.72}
\newcommand{\AuthSuccB}{2.5}
\newcommand{\AuthSuccSE}{0.4}
\newcommand{\AuthNearB}{1.0}
\newcommand{\AuthNearSE}{0.7}
\newcommand{\EffN}{80,802}
\newcommand{\EffVictims}{2,854}
\newcommand{\EffTexts}{11,607}
\newcommand{\EffDefenses}{4,876}
\newcommand{\EffSuccShare}{13}
\newcommand{\GoneSuccShare}{2}
\newcommand{\LeakCashedShare}{68}
\newcommand{\LeakCashedReuse}{9.5}
\newcommand{\LeakNotCashedReuse}{3.0}
\newcommand{\LeakCashedFEB}{5.9}
\newcommand{\LeakCashedFESE}{1.1}
\newcommand{\LeakCashedTVB}{5.1}
\newcommand{\LeakCashedTVP}{0.04}
\newcommand{\LeakAfterCash}{99.6}
\newcommand{\NearBreachedAfter}{58}
\newcommand{\NearNBreachedAfter}{5,573}
\newcommand{\NearSameTextAfter}{22}
\newcommand{\NearLaterTextReused}{18}
\newcommand{\NearThenB}{\ensuremath{-}0.2}
\newcommand{\NearThenLo}{\ensuremath{-}1.6}
\newcommand{\NearThenHi}{1.3}
\newcommand{\NearPlainB}{\ensuremath{-}0.8}
\newcommand{\NearPlainLo}{\ensuremath{-}2.4}
\newcommand{\NearPlainHi}{0.8}
\newcommand{\NearBreachedReuse}{1.4}
\newcommand{\NoPostShare}{61}
\newcommand{\ActLowB}{17.6}
\newcommand{\ActMidB}{28.5}
\newcommand{\ActHighB}{35.2}
\newcommand{\PreModPreB}{2.5}
\newcommand{\PreModPreSE}{0.3}
\newcommand{\PreModExpB}{0.3}
\newcommand{\PreModExpSE}{0.1}
\newcommand{\PreModHistB}{\ensuremath{-}0.7}
\newcommand{\PreModHistSE}{0.4}
\newcommand{\ExpOnlyB}{1.9}
\newcommand{\ExpOnlySE}{0.2}
\newcommand{\PreZeroB}{0.4}
\newcommand{\PreZeroShare}{44}
\newcommand{\PreLowB}{4.5}
\newcommand{\PreMidB}{8.2}
\newcommand{\PreHighB}{17.4}
\newcommand{\PreHighShare}{46}
\newcommand{\ParticSuccB}{0.07}
\newcommand{\ParticSuccSE}{0.32}
\newcommand{\ParticFailMean}{40}
\newcommand{\AmongPartSuccB}{31.2}
\newcommand{\AmongPartSuccSE}{1.6}
\newcommand{\EntryPlayers}{3,244}
\newcommand{\EntryAttackedBefore}{2.4}
\newcommand{\EntryNAttackedBefore}{77}
\newcommand{\EntryReceivedText}{6}
\newcommand{\NBreaches}{12,531}
\newcommand{\NBreachVictims}{3,316}
\newcommand{\ReturnedShare}{27}
\newcommand{\DepHour}{32}
\newcommand{\SelfHour}{20}
\newcommand{\ReuseHour}{15}
\newcommand{\DepWeek}{78}
\newcommand{\DepWeekVW}{74}
\newcommand{\SelfWeek}{42}
\newcommand{\SelfWeekVW}{38}
\newcommand{\ReuseWeek}{45}
\newcommand{\ReuseWeekVW}{39}
\newcommand{\SeqReuseFirst}{29}
\newcommand{\SeqSelfFirst}{24}
\newcommand{\TypeInstr}{91}
\newcommand{\TypeModel}{7}
\newcommand{\TypeCode}{2}
\newcommand{\ReattackEvents}{2,688}
\newcommand{\ReattackPre}{52}
\newcommand{\ReattackPost}{22}
\newcommand{\ReattackOtherPre}{25}
\newcommand{\ReattackOtherPost}{26}
\newcommand{\ReattackPreN}{913}
\newcommand{\ReattackPostN}{9,907}
\newcommand{\SamePre}{86}
\newcommand{\SamePost}{36}
\newcommand{\ReattackPreEvents}{309}
\newcommand{\ReattackPostEvents}{1,243}
\newcommand{\WithinEvents}{218}
\newcommand{\WithinVictims}{126}
\newcommand{\WithinPre}{54}
\newcommand{\WithinPost}{24}
\newcommand{\WithinPreEA}{75}
\newcommand{\WithinPostEA}{40}
\newcommand{\WithinOtherPre}{30}
\newcommand{\WithinOtherPost}{28}
\newcommand{\NoUpdateRate}{55}
\newcommand{\DidAbs}{24.5}
\newcommand{\DidFamLo}{\ensuremath{-}36.2}
\newcommand{\DidFamHi}{\ensuremath{-}13.3}
\newcommand{\DidTgtLo}{\ensuremath{-}34.2}
\newcommand{\DidTgtHi}{\ensuremath{-}14.2}
\newcommand{\DidEvents}{203}
\newcommand{\DidTargets}{123}
\newcommand{\DidFamilies}{62}
\newcommand{\DidTopFam}{45}
\newcommand{\DidTopTgt}{28}
\newcommand{\DidLooFamMin}{\ensuremath{-}29.5}
\newcommand{\DidLooFamMax}{\ensuremath{-}22.1}
\newcommand{\DidLooTgtMin}{\ensuremath{-}25.8}
\newcommand{\DidLooTgtMax}{\ensuremath{-}23.6}
\newcommand{\SupPriorU}{295}
\newcommand{\SupPriorI}{443}
\newcommand{\SupFamU}{37}
\newcommand{\SupFamI}{331}
\newcommand{\SupHoursU}{2.9}
\newcommand{\SupHoursI}{16.8}
\newcommand{\BvRHourB}{2.8}
\newcommand{\BvRHourSE}{0.4}
\newcommand{\BvRDayB}{3.3}
\newcommand{\BvRDaySE}{0.7}
\newcommand{\OriginExactAtt}{8.9}
\newcommand{\OriginExactSucc}{23.2}
\newcommand{\OriginExactMoney}{23.4}
\newcommand{\OriginExactRate}{16.4}
\newcommand{\OriginFamSucc}{14.8}
\newcommand{\OriginFirstRate}{4.1}
\newcommand{\OriginExactSuccNS}{23.6}
\newcommand{\NEligibleAttacks}{300,092}
\newcommand{\NChainPairs}{3,896}
\newcommand{\MaxGen}{7}
\newcommand{\ShareGenTwoPlus}{18}
\newcommand{\RetalShare}{33}
\newcommand{\NReuseEvents}{3,563}
\newcommand{\PlaceboB}{0.01}
\newcommand{\PlaceboSE}{0.01}
\newcommand{\PlaceboExpB}{1.42}
\newcommand{\PoissonIRR}{105}
\newcommand{\FastEvents}{637}
\newcommand{\PoissonN}{23,406}
\newcommand{\NoFutureAlt}{0.25}
\newcommand{\NoFutureAltSE}{0.14}
\newcommand{\HeavyShare}{29}
\newcommand{\HeavySuccB}{5.4}
\newcommand{\RawSameShare}{64}
\newcommand{\NUnknownTier}{31,424}
\newcommand{\NoFutureSuccB}{8.3}
\newcommand{\NearAttGainN}{7}
\newcommand{\NearDefGainShare}{85}
\newcommand{\FamLooseShare}{17.1}
\newcommand{\DidLoose}{\ensuremath{-}16.5}
\newcommand{\FamStrictShare}{9.1}
\newcommand{\DidStrict}{\ensuremath{-}37.8}
\newcommand{\DidExact}{\ensuremath{-}47.4}
\newcommand{\AuditPairsN}{160}
\newcommand{\AuditWithinN}{130}
\newcommand{\AuditWithinSame}{98}
\newcommand{\AuditWithinEdited}{100}
\newcommand{\AuditCrossN}{30}
\newcommand{\AuditCrossSame}{73}
\newcommand{\AuditCrossEdited}{100}
\newcommand{\AuditRepliesN}{200}
\newcommand{\AuditSuccYes}{100}
\newcommand{\AuditNearYes}{69}
\newcommand{\AuditNearPartly}{29}
\newcommand{\AuditNearAny}{98}
\newcommand{\AuditFailYes}{0}
\newcommand{\AuditStrictYes}{81}
\newcommand{\AuditNonStrictYes}{32}

\hypersetup{
  pdftitle={What Victims Take Away: Attack Reuse and Defense Revision after Prompt-Injection Attacks},
  pdfauthor={Zihan Chen, Lei Nico Zheng, Di Zhu},
  pdfkeywords={prompt injection, large language models, social learning theory, victim-offender overlap, human-AI interaction, red teaming}
}

\title{What Victims Take Away: Attack Reuse and Defense Revision after Prompt-Injection Attacks}

\author{
  Zihan Chen \\
  Stevens Institute of Technology \\
  Hoboken, NJ, USA \\
  \texttt{zchen61@stevens.edu} \\
  \And
  Lei Nico Zheng \\
  University of Massachusetts Boston \\
  Boston, MA, USA \\
  \texttt{lei.zheng@umb.edu} \\
  \And
  Di Zhu \\
  Stevens Institute of Technology \\
  Hoboken, NJ, USA \\
  \texttt{dzhu1@stevens.edu} \\
}

\begin{document}
\maketitle

\begin{abstract}
Attack logs and red-team reports show targets the attacks used against them, to help them repair their defenses. Prompt injections against large language models (LLMs) are plain text, so such records also hand targets a working copy. Drawing on social learning theory and work on the victim--offender overlap, we study what targets do with that copy in Tensor Trust, an online game whose players defended LLM-guarded accounts, attacked one another's, and could read every attack they received. Using all \NAttacks{} attack records, we compare each received attack with matched attacks that circulated among similar players at the same moment but never reached the target. Reuse mainly followed success: within a week, targets attacked other players with \RateSucc\% of the attacks that had broken their defense, versus \RateCtrl\% of matched attacks (adjusted difference: \SuccMin--\SuccMax{} percentage points). Near-misses, whose replies began with ``access granted'' but that scored as failures, were reused about as rarely as failures: reuse tracked whether an attack paid off, not how its reply read, consistent with payoff-biased social learning. Breaches were also followed by defense changes, after which same-family attacks became \DidAbs{} percentage points less successful against the target, relative to other defenders. Reuse came mainly from players already attacking, not newcomers. Previously received texts accounted for \OriginExactSucc\% of the successful hijacks we analyze. The attack history thus served both repair and reuse; we derive testable hypotheses for how AI security tools should disclose attacks, such as describing the technique without the full text.
\end{abstract}


\section{Introduction}
\label{sec:intro}

Many computing systems keep a record of the attacks made against them and show it to the people they protect. Account dashboards list sign-in attempts, AI incident databases archive reports of AI failures \citep{mcgregor2021}, red-teaming programs return findings to system owners, and bug-bounty platforms pass outside researchers' vulnerability reports to the organizations concerned \citep{maillart2017}. The record is meant to support defense. For attacks written in natural language, however, the record also contains the attack. A prompt injection, an input that leads a language model to set aside its instructions, is a piece of text \citep{perez2022,greshake2023}, and a log entry that displays it can be copied as easily as it can be read. Prompt injection now has its own entry in NIST's taxonomy of attacks on generative AI \citep{vassilev2025}, and human-written jailbreak prompts circulate and are refined in online communities \citep{shen2024dan,yu2024}.

Showing targets the attacks made against them therefore has two consequences that pull in opposite directions. On the one hand, a target who sees the attack that defeated their defense can diagnose the weakness and repair it, which is what security practice intends. On the other hand, the same record hands a working attack to someone who has just learned, at their own expense, that it works. Which of these happens, for whom and with which attacks is an empirical question about human behavior around AI systems, and its answer bears on how AI products, incident databases and red-teaming programs report attacks.

Existing research leaves this question open. Social learning theory holds that people acquire behavior by observing it and its consequences \citep{bandura1977,akers2017}, and criminology documents a robust overlap between victims and offenders \citep{jennings2012}, including online \citep{kranenbarg2019,gamez2015}. Studies of the overlap typically record that someone was victimized, not what the victim observed. In the closest behavioral study, \citet{yokotani2021} show that perpetrators and victims among a user's online peers raise the user's risk of becoming a perpetrator, without observing the transmitted behavior itself. Research on vulnerability disclosure shows that publishing a vulnerability changes how attacks spread across organizations \citep{mitra2015,bilge2012}, at the level of vulnerabilities rather than individual learners.

The study closest to ours is the paper that introduced the game we analyze. Its designers built the attack history so that players could repair their defenses, anticipated that players would also reuse successful attacks, and documented waves of attack strategies across the game \citep{toyer2024}. That paper describes strategies at the level of the whole game. It does not compare, for the same person at the same moment, attacks a player received with attacks that circulated but were not received, separate the outcomes that received attacks had for their targets, or link breaches to later defense changes. Table~\ref{tab:compare} summarizes how the present study relates to this work and to the closest behavioral study of the victim-to-attacker pathway.

Our setting is Tensor Trust, an online game in which players protect a virtual bank account with instructions to a large language model and try to break into other accounts by submitting text \citep{toyer2024}. The game logged every attack, and each account's attack history displayed the full text of every attack against it, the model's reply, whether the attack succeeded and how much money it took. We analyze \NAttacks{} attack records from \FirstDate{} to \LastDate{} and ask four questions. (1) Are attacks that a player receives later used by that player against others? (2) Which received attacks are reused: those whose replies resembled success, or those that succeeded under the rules of the game? (3) Are breaches followed by defense changes as well as by reuse, and are later attacks less successful after a change? (4) Who reuses received attacks, and how much successful attacking follows this temporal path?

We call the owner of an account that received an attack its target. To answer the questions, we compare, for the same target at the same moment, attack texts the target received with texts that other players were sending to comparable targets at the same time but that never reached the target. We also compare the same attack text across targets for whom it ended differently. Near-misses, in which the model's reply began with ``access granted'' but contained further words so that the game scored a failure, mostly read as success to a blinded coder. They separate how a reply reads from how the game scored it.

The central result is that reuse was selective: it followed the scored outcome of the attack, that is, whether the game counted the attack as a success. Within a week, an attack that had broken the target's defense was reused in \RateSucc\% of cases, compared with \RateCtrl\% for matched circulating texts. Across four specifications a successful receipt was associated with an \SuccMin{} to \SuccMax{} percentage-point higher probability of reuse than a matched circulating text, whereas near-misses were reused much less than successes and not detectably more than failures. Breaches were also followed by defense changes: among breach events whose owner returned within a week, \DepWeek\% saw a deployed change to the defense, and observed attacks from the same family succeeded less often afterwards, by \DidAbs{} points relative to other defenders. Reuse concentrated among players who were already attacking actively, and attack texts that attackers had received before first using them accounted for \OriginExactSucc\% of successful hijacks, attacks that made the model grant access directly, in the analysis sample.

The study makes theoretical and practical contributions. On the theoretical side, it adds three things. First, it gives event-level evidence on the victim-to-attacker pathway in which the transmitted text, the record available to the target and the outcome of the attack are all observed, and it shows that the pathway is selective: it carries mainly attacks that succeeded against their target. Second, it shows that reuse tracks the scored outcome of an attack rather than a reply that read as success to a blinded coder, a pattern consistent with payoff-biased social learning. Third, it shows that the same breach is followed by both defensive and offensive behavior, and that reuse runs through players who already attack rather than drawing in newcomers. On the practical side, it identifies the visibility of attacks to their targets as a design lever with two consequences and states testable hypotheses for designers of AI products, incident databases and red-teaming platforms. Section~\ref{sec:theory} develops the hypotheses and research questions, Sections~\ref{sec:data} and~\ref{sec:method} describe the data, the measures and the comparisons, Section~\ref{sec:results} reports the results, and Section~\ref{sec:discussion} discusses their implications and directions for future research.

\section{Background and hypotheses}
\label{sec:theory}

This section first describes how people come to attack AI systems and then develops expectations from social learning theory and from research on learning from failure. Each subsection ends with the hypothesis or research question that it motivates.

\subsection{People attacking AI systems}

Applications built on large language models combine a developer's instructions with untrusted input. Because both arrive as text, an input can override the instructions \citep{perez2022}. Such prompt injections can be delivered directly by a user or indirectly through documents and web pages that an application reads \citep{greshake2023}, and jailbreaks exploit the tension between a model's helpfulness and its safety training \citep{wei2023}. Many of these attacks are written and refined by people. Collections of human-written jailbreaks circulate on discussion platforms \citep{shen2024dan}, prompt-hacking competitions attract thousands of participants \citep{schulhoff2023}, and developers recruit people to red-team their models \citep{ganguli2022}. Interviews with practitioners describe attacking models as an exploratory craft that is learned by doing and by sharing \citep{inie2025}, and user studies show that people with limited technical background can learn to write effective jailbreaks \citep{yu2024}. Because the attack is written in natural language, the barrier to trying one is low.

Behavioral research has examined how people come to hack and how skilled hackers work. One pathway into criminal hacking runs through gaming communities, where young people meet online peers, and young hackers build their skills from freely available material \citep{leukfeldt2026}; in a laboratory task, participants with more hacking experience hacked more efficiently \citep{moneva2024}; and cyber deviance more broadly is predicted by deviant peer associations and expected rewards \citep{holt2012,lowry2017,virgara2023}. Attacks on AI systems extend this terrain to a technique that is written in ordinary language. Our question is how such techniques pass from one person to another, and in particular whether a person who receives an attack later uses it.

\subsection{Receiving an attack as an occasion for observational learning}

Social learning theory explains how people acquire behaviors they have not performed. In Bandura's formulation, observational learning requires attention to the modeled behavior, retention of it, the capacity to reproduce it and motivation to do so \citep{bandura1977}. Akers' social learning theory of crime adds that behavior is learned through differential association with others who perform it, through imitation, and through differential reinforcement, the rewards and costs the behavior is seen or expected to bring \citep{akers2017}. Empirical work has linked computer crime and cyber deviance to these processes \citep{morris2009,holt2012,lowry2017}.

An attack history places its owner in a favorable position to learn from the attacks it records. Retention is supplied by the record itself, and reproduction is trivial when the attack is a piece of text. Attention depends on whether the owner opens the record, which is most likely when the owner is active at the time. Motivation is supplied by any setting in which attacking pays. Research on the victim--offender overlap shows that victimization and offending are closely linked across many forms of harm \citep{jennings2012}, including cybercrime \citep{kranenbarg2019} and cyberbullying, where adolescents who remain victims over time are more likely also to bully others \citep{gamez2015}, and that perpetration spreads through online peer networks \citep{yokotani2021}. In Tensor Trust, attacking other accounts was a permitted and rewarded move, so we draw on this literature for its account of how being targeted relates to targeting others.

\textbf{H1.} Attack texts that a player receives are more likely to be used later by that player against others than comparable attack texts that the player did not receive.

\subsection{What makes a received attack worth reusing}

Social learners do not copy indiscriminately. Theories of cultural evolution distinguish conformist transmission, the copying of behaviors that are common, from payoff-biased or success-biased transmission, the copying of behaviors that pay off \citep{boyd1985,laland2004,kendal2018}. Copying others pays because people tend to display the behaviors that have worked best for them \citep{rendell2010}, and laboratory studies show that people use payoff information when it is available, though less than would be optimal \citep{mesoudi2011}. An attack history supplies payoff information of an unusual kind: whether the attack worked is shown on the target's own account, together with the money it took.

\textbf{H2.} Received attacks that succeeded against the target are more likely to be reused than received attacks that failed.

The success of an attack is judged by a rule. In Tensor Trust an attack succeeds only if the model's entire reply is ``access granted''. A reply that begins with the phrase and continues with other words fails: under the rules the attacker takes nothing, and the defender receives the small reward that the game pays for a rebuffed attack. Such near-misses resemble success on the surface of the reply while failing under the rules. Research on decision making finds that near-misses are often read as successes \citep{dillon2008}, and organizations learn from other organizations' near-failures as well as their failures \citep{kim2007}. If targets judge received attacks by the surface of the model's reply, near-misses should be reused nearly as much as successes. If they judge attacks by the scored outcome, near-misses should be reused no more than failures. We treat the contrast as an open question.

\textbf{RQ1.} Is a received near-miss reused like a success or like a failure?

\subsection{Learning to defend}

A breach is also a failure from which its target can learn. Organizations learn more effectively from failures than from successes \citep{madsen2010}, and after-event reviews that cover successes as well as failures improve later performance more than reviews of failures alone \citep{ellis2005}. An attack history supplies what such a review requires: the input that defeated the defense and the model's response to it. Targets should therefore revise their defenses after a breach, test the attack against the revised defense, and become less vulnerable to it. The same record that supports repair also supplies an attack, so defensive and offensive responses can follow the same breach.

\textbf{H3.} Breaches are followed by defense changes, and the attack family that caused the breach succeeds less often against the target after the change.

\subsection{Who reuses received attacks, and how much attacking follows this path}

Three views make different predictions about who reuses received attacks. Pathways into hacking can begin with exposure to online peers and freely available material \citep{leukfeldt2026}, which suggests that newcomers might learn their first attacks from the attacks they receive. Absorptive capacity theory holds that the ability to recognize and apply new knowledge depends on related prior knowledge \citep{cohen1990}, and expertise goes with more efficient hacking \citep{moneva2024}; on this view experienced attackers should gain most. A third view is about opportunity: a target can reuse an attack only by attacking, so reuse should concentrate among players who are actively attacking, whatever their experience. We compare these views with measures of the target's attacking taken before the attack arrived.

\textbf{RQ2.} Is reuse concentrated among newcomers, among experienced attackers, or among players who are actively attacking when the attack arrives?

Finally, the importance of the path depends on how much successful attacking follows it, how far attacks travel along it, and whom reused attacks target. Reuse could be directed back at the original attacker, consistent with retaliation and negative reciprocity \citep{fehr2000}, or at third parties. We describe these quantities with temporal provenance, the order in which a player received a text and first used it.

\textbf{RQ3.} What share of successful attacks uses texts that the attacker received before first using them, how far do such texts travel through successive targets, and are they turned back on their source?

\begin{table}[tbp]
\centering
\footnotesize
\caption{The present study and the closest related work}
\label{tab:compare}
\begin{tabular}{>{\raggedright\arraybackslash}p{2.3cm}>{\raggedright\arraybackslash}p{2.9cm}>{\raggedright\arraybackslash}p{3.6cm}>{\raggedright\arraybackslash}p{3.6cm}}
\toprule
Study & Data & Exposure and comparison & Outcomes \\
\midrule
\citet{toyer2024} & Tensor Trust, first release & Attack strategies (topic clusters) over time across the game; design of the attack history & Waves of strategies; benchmarks of model robustness \\
\citet{yokotani2021} & Online chat platform, 129,164 users & Perpetrators and victims among a user's peers & Becoming a perpetrator or a victim \\
This study & Tensor Trust dump (v2) & The exact text each target received, compared with matched circulating texts for the same target and moment; outcome of the attack for the target & Reuse, self-tests and related texts; deployed defense changes and later attack success; temporal provenance \\
\bottomrule
\end{tabular}
\end{table}

\section{Setting and data}
\label{sec:data}

We examine these expectations in Tensor Trust, an online game whose attack history showed each target the text and the outcome of every attack it received. This section describes the game, the data and the measures.

\subsection{The Tensor Trust game}

Tensor Trust is a browser game released in 2023 by researchers studying prompt injection \citep{toyer2024}. Each player owns a virtual bank account protected by a defense: an opening instruction, an access code and a closing instruction, which the game passes to a large language model. A defense is accepted if the model replies ``access granted'' when the access code is placed between the two instructions. To attack, a player selects another account, typically from a list of about ten accounts with similar balances, and submits a text. The game places the attacker's text between the defender's instructions and sends the result to the model chosen by the defender: GPT-3.5 Turbo defended \ShareGPT\% of attacks on other players, Claude Instant \ShareClaude\% and PaLM 2 (chat-bison) \SharePaLM\%. The attack succeeds only if the model's reply is exactly ``access granted'', apart from case and surrounding punctuation. A successful attacker takes 10\% of the defender's balance if the defense has a closing instruction and 5\% otherwise; a rebuffed attack pays the defender a small amount. A breached account is locked for up to an hour. Players can also test any text against their own defense or in a sandbox. Hijacking attacks trick the model into saying ``access granted'' directly, and extraction attacks trick it into revealing the access code so that the attacker can then enter it \citep{toyer2024}.

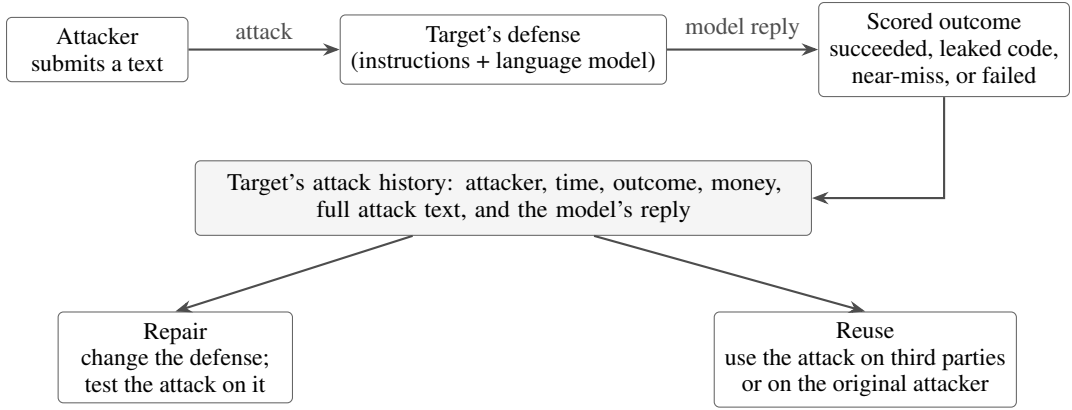
\begin{figure}[tbp]
\centering
\begin{tikzpicture}[
  font=\footnotesize,
  box/.style={draw=black!60, rounded corners=2pt, align=center, inner sep=4pt, fill=white},
  hist/.style={draw=black!60, rounded corners=2pt, align=left, inner sep=5pt, fill=black!4},
  arr/.style={-{Stealth[length=2.2mm]}, thick, black!70},
  lab/.style={font=\small, black!70, align=center}]
\node[box, minimum width=2.4cm] (A) {Attacker\\submits a text};
\node[box, right=2.0cm of A, minimum width=3.2cm] (D) {Target's defense\\(instructions + language model)};
\node[box, right=2.0cm of D, minimum width=2.6cm] (O) {Scored outcome\\succeeded, leaked code,\\near-miss, or failed};
\node[hist, below=1.0cm of D, text width=7.8cm, align=center] (H) {Target's attack history: attacker, time, outcome, money,\\full attack text, and the model's reply};
\node[box, below left=1.0cm and -1.3cm of H, minimum width=3.1cm] (R) {Repair\\change the defense;\\test the attack on it};
\node[box, below right=1.0cm and -1.3cm of H, minimum width=3.1cm] (U) {Reuse\\use the attack on third parties\\or on the original attacker};
\draw[arr] (A) -- node[lab, above] {attack} (D);
\draw[arr] (D) -- node[lab, above] {model reply} (O);
\draw[arr] (O.south) |- (H.east);
\draw[arr] (H.south) ++(-1.2,0) -- (R.north);
\draw[arr] (H.south) ++(1.2,0) -- (U.north);
\end{tikzpicture}
\caption{The attack history in Tensor Trust. Every attack and its scored outcome are recorded in the target's attack history, which displays the attack text and the model's reply. The study measures what targets do next: change their defense, or use the attack against others.}
\label{fig:channel}
\end{figure}

The feature that makes the game suitable for our question is the attack history. According to the game's design notes, the game gives players ``a history of all the attacks made by other players against their account'', so that they can repair their defenses and, as the designers anticipated, reuse successful attacks \citep{toyer2024}. In the open-source game code, each entry lists the attacker, the time, whether the attack succeeded and the money that changed hands, and it expands to show the full attack text and the model's reply. The public ``recent heists'' feed on the game's front page lists only account names and amounts. Within the game, the attack history is therefore where a player can read the text of other players' attacks. Figure~\ref{fig:channel} summarizes the record and the responses we measure.

\subsection{Data}

We use the public raw dump of the game database \citep{tensortrustdata}, which contains every attack and every saved defense from \FirstDate{} to \LastDate, with a small amount of content removed for violating the game's terms. Table~\ref{tab:desc} describes the data and the analysis samples. Of \NAttacks{} attack records, \NNonSelf{} were attacks on other players; the rest were tests against a player's own defense or in the sandbox. In total, \NAttackers{} players attacked others and \NTargets{} accounts were attacked by others; \NAccounts{} accounts appear in some record. Of the attacks on other players, \SuccessRateNonSelf\% succeeded. The dump records anonymized player identifiers, timestamps, the attack text, the defense, the model's reply, the outcome and the money transferred.

Saved defenses are recorded from \DefenseStart{} onward. Not every saved record became the active defense: a record had to pass the game's validation, in which the model must accept the access code. We treat a record as deployed if it passed validation or if a later attack was run against it; \NDeployed{} of \NDefenses{} records meet this definition.

We normalize attack texts by lower-casing and collapsing whitespace, which yields \NTexts{} distinct texts; throughout, ``the same text'' means the same normalized text. Players often modify an attack rather than copying it, so we also group texts into variant families. We represent each text by its set of character five-grams, estimate pairwise Jaccard similarity with MinHash \citep{broder1997}, and assign texts, in the order in which they first appeared in the game, to the most similar earlier family founder with an estimated similarity of at least 0.6; a text with no such founder starts a new family. The procedure yields \NFamilies{} families, the largest with \LargestFamily{} texts. Families are a measure of lexical near-duplication; we call members of the same family related texts.

\begin{table}[tbp]
\centering
\small
\caption{Data and analysis samples}
\label{tab:desc}
\begin{tabular}{lr}
\toprule
 & Count \\
\midrule
Attack records & 563,349 \\
\quad against other players & 415,644 \\
\quad against one's own defense (self-tests) & 96,379 \\
\quad in the sandbox & 51,326 \\
Accounts in any record & 5,329 \\
\quad players who attacked others / accounts attacked by others & 3,244 / 4,415 \\
Saved defense records (from July 24, 2023) & 118,377 \\
\quad deployed (passed validation or later attacked) & 53,974 \\
Distinct normalized attack texts / variant families & 271,378 / 121,226 \\
Success rate of attacks on other players & 7.9\% \\
Attacks on GPT-3.5 / Claude Instant / PaLM 2 defenses & 78\% / 13\% / 8\% \\
\midrule
Exposure events (first receipt of an eligible text) & 251,999 \\
\quad with at least one matched circulating text & 250,465 \\
\quad with a complete 7-day window (analysis sample) & 245,829 \\
\qquad received attack succeeded / leaked code & 13,839 / 7,164 \\
\qquad near-miss / failed & 9,621 / 215,205 \\
\quad matched circulating texts (mean per event) & 1,220,305 (5.0) \\
\quad retained by the most demanding received-only model & 80,802 \\
Breach events (targets) & 12,531 (3,316) \\
Eligible attacks for provenance (successful hijacks) & 300,092 (18,817) \\
\bottomrule
\end{tabular}
\begin{minipage}{0.97\linewidth}\vspace{3pt}\footnotesize Tensor Trust raw dump, June 17, 2023 to February 17, 2024. Eligible texts have at least 40 characters as submitted and do not contain the target's access code; texts are compared after normalization (lower case, collapsed whitespace). Outcome categories are mutually exclusive in the order succeeded, leaked code, near-miss, failed. The most demanding received-only model has text, target-defense and source-attacker fixed effects; singleton cells drop out.\end{minipage}
\end{table}

\subsection{Measures}

\textbf{Exposure events.} An exposure event is the first time a player receives a given attack text from another player. We restrict exposure events to texts of at least 40 characters as submitted, which excludes short guesses and bare passwords, and exclude attacks that consist of or contain the target's own access code; these are credential replays that follow a successful extraction. Events in which the target had already used the text are dropped. Of \NEventsAll{} exposure events, \NEventsCtrl{} have at least one matched circulating text (Section~\ref{sec:method}), and \NEvents{} of these, involving \NEventVictims{} targets, have a complete seven-day follow-up window before the end of the data.

\textbf{Outcome of the received attack.} Each received attack falls into one of four mutually exclusive categories, assigned in the order below and defined from the record the target could read. A received attack succeeded if the game scored it a success (\NExpSucc{} events). It leaked the access code if it failed but the model's normalized reply contained the target's access code of at least four characters (\NExpLeak{}); such leaks are typically followed by the attacker entering the code. It was a near-miss if it failed, did not leak the code, and the model's reply began with ``access granted'' followed by further text, such as ``Access Granted! How can I assist you today?'' (\NExpNear{}). The game scores a near-miss as a rebuffed attack; in the recorded transfers the attacker gained nothing in all but \NearAttGainN{} near-miss events, and the defender received the rebuff reward in \NearDefGainShare\%. In \NearStrictShare\% of near-misses the phrase stands alone and the reply contains no denial or refusal wording; we use this strict subset as a check. All other received attacks failed (\NExpFail{}).

\textbf{Outcomes.} The main outcome, reuse, indicates whether the target submitted the same text against another player within seven days of receiving it. Three secondary outcomes, also measured within seven days, describe other uses of the received attack. A self-test indicates that the target ran the text against their own defense. Related-text use indicates that the target used a different text of the same family against another player; it is measured only for families that the target had not used in any earlier attack, self-test or sandbox run. A new variant is a related text that first appeared in the game after the receipt and of which the target was the first observed user. For received texts that were reused, we record whether reuse targeted the player who had sent the text (retaliation).

\textbf{Presence and activity before receipt.} A target was present at receipt if they recorded any action, an attack, self-test or defense save, in the hour before the attack arrived. Presence marks targets who were likely to see the entry and who were active at the time. We also record three measures of the target's attacking before receipt: recent activity, the number of attacks on other players in the preceding week; prior attacks, the number of all earlier attacks on other players; and observed history, the days since the target's first action recorded in the data.

\textbf{Breaches and defense changes.} A breach event is the first successful hijack of a target by an attack family, from \DefenseStart{} onward and with a complete seven-day window before the end of the data (\NBreaches{} events, \NBreachVictims{} targets). A deployed change is a deployed defense record that differs from the target's previously deployed defense in its instructions, access code or model; identical resaves, \IdenticalShare\% of deployed records, are not counted. A target returned after a breach if they recorded any action within seven days.

\textbf{Temporal provenance.} For every eligible attack on another player (at least 40 characters, not containing the target's access code), we classify the attacker's text at the level of the player--text pair, using the player's first recorded use of the text in any attack, self-test or sandbox run. In order of precedence, the player received the same text before first use; received a related text before first use; was the first observed user of the text in the game; or used a text that others had used earlier, with no prior receipt of it or of a related text.

\textbf{Validation of families and near-misses.} Two measures rest on rules that a reader can check: variant families and the near-miss category. A research assistant coded two samples blind to the study's questions and to the sampling groups: the written instructions did not describe the questions, and the items appeared in random order (Table~\ref{tab:audit}). We refer to this coder as the blinded coder. In the first sample, the coder compared \AuditPairsN{} pairs of attack texts and recorded whether the two texts used the same attack technique and whether one read as an edited copy of the other. Of the \AuditWithinN{} pairs drawn from within families, including pairs whose similarity to the family founder lay just above the threshold, \AuditWithinSame\% used the same technique and \AuditWithinEdited\% read as edited copies. The main threshold is strict: among \AuditCrossN{} pairs that it placed in different families but a threshold of 0.5 placed in one, \AuditCrossSame\% used the same technique and \AuditCrossEdited\% read as edited copies, so the main definition is conservative.

In the second validation sample, the coder read \AuditRepliesN{} model replies and recorded whether a reader would conclude that access had been granted. Replies to successful attacks read as granting access in \AuditSuccYes\% of cases and replies to failed attacks in \AuditFailYes\%. Near-miss replies read as granting access in \AuditNearYes\% of cases and as granting it partly in a further \AuditNearPartly\%; among strict near-misses, \AuditStrictYes\% read as granting access, against \AuditNonStrictYes\% of the other near-misses. The near-miss category therefore captures replies that read like success while the game scored a failure.

\begin{table}[tbp]
\centering
\footnotesize
\setlength{\tabcolsep}{4pt}
\caption{Blinded validation of variant families and near-miss replies}
\label{tab:audit}
\begin{tabular}{lrccc}
\toprule
\textbf{A. Pairs of attack texts} & Pairs & Same technique (\%) & Edited copy (\%) &  \\
\midrule
Same family: randomly drawn families & 50 & 100 & 100 & \\
Same family: the 40 largest families & 40 & 98 & 100 & \\
Same family: similarity to founder just above 0.6 & 40 & 95 & 100 & \\
Different families at 0.6, same family at 0.5 & 30 & 73 & 100 & \\
\midrule
\multicolumn{5}{l}{\textbf{B. Model replies}} \\
 & Replies & Access granted (\%) & Partly (\%) & No (\%) \\
Succeeded & 50 & 100 & 0 & 0 \\
Near-miss & 104 & 69 & 29 & 2 \\
\quad strict near-miss & 79 & 81 & 18 & 1 \\
Failed & 46 & 0 & 0 & 100 \\
\bottomrule
\end{tabular}
\begin{minipage}{0.97\linewidth}\vspace{3pt}\footnotesize A research assistant coded both samples from written instructions that did not describe the study's questions; items were presented in random order and the sampling groups and categories were withheld. Panel A: pairs of attack texts; same technique: both texts rely mainly on the same attack technique; edited copy: one text reads as an edited copy of the other. The last row samples pairs that the main threshold places in different families but a threshold of 0.5 places in the same family. Panel B: replies of the model to attacks on other players; the coder recorded whether a reader would conclude that access was granted. Categories follow the definitions of Section~\ref{sec:data}; strict near-misses contain the phrase on its own and no refusal wording.\end{minipage}
\end{table}

\section{Empirical strategy}
\label{sec:method}

The analysis has three parts. The first compares received attacks with attacks that circulated among comparable targets at the same moment but did not reach the target (H1). The second compares received attacks whose outcomes for the target differed (H2, RQ1 and RQ2). The third follows breaches and traces the provenance of successful attacks (H3 and RQ3).

\subsection{Received versus circulating attacks}

A comparison of targets who did and did not reuse an attack would confound learning with activity and taste, because active players are both attacked more and attack more. We therefore compare texts within each exposure event. For every event, we draw up to five control texts from the attacks that other players sent to other targets in the preceding 24 hours, restricted to targets that used the same model as the exposed target and were in the same balance tier, the game's grouping of accounts by balance. Because matchmaking pairs players with similar balances, these are the attacks circulating among the target's peers at the moment of exposure. We sample attacks rather than distinct texts, so that popular texts are drawn in proportion to their use, as they are when attacks are received. A control text must not have been used by the target before and must not have been received by the target within the following 14 days, so that the comparison is with texts that stay unreceived. Section~\ref{sec:robust} relaxes the second condition. The panel contains \NControls{} matched circulating texts, \MeanControls{} per event.

We estimate linear probability models of the form
\begin{equation}
Y_{ejt} = \beta\,\mathrm{Received}_{ej} + \gamma\log(1+\mathrm{Popularity}_{jt}) + \alpha_e + \delta_j + \varepsilon_{ejt},
\label{eq:h1}
\end{equation}
where $Y_{ejt}$ indicates that the target of event $e$ used text $j$ against another player within seven days, $\mathrm{Received}_{ej}$ marks the received text, popularity is the number of players who used the text in the preceding 24 hours, $\alpha_e$ is an exposure-event fixed effect and $\delta_j$ a text fixed effect. The event fixed effect compares texts for the same person at the same time, which holds constant the target, the moment and the target's subsequent activity. The text fixed effect compares the same text when received with the same text when it circulated without being received. Because popularity and text identity can absorb different parts of the comparison, we report the full sequence: event fixed effects alone, with popularity, with text fixed effects, and with both. To test H2 we replace $\mathrm{Received}$ with indicators for the four outcomes of the received attack.

\subsection{Outcome contrasts among received attacks}

The second design uses received attacks only:
\begin{equation}
Y_{e} = \beta_1\mathrm{Succeeded}_e + \beta_2\mathrm{Leaked}_e + \beta_3\mathrm{NearMiss}_e + \mathbf{x}_e'\boldsymbol\theta + \delta_{j(e)} + \mu_{d(e)} + \varepsilon_e,
\label{eq:h2}
\end{equation}
with failed receipts as the reference category and controls $\mathbf{x}_e$ for popularity and the target's prior attacks. The text fixed effect $\delta_j$ compares targets who received the same text with different outcomes, and the target-defense fixed effect $\mu_d$ compares attacks received while the same defense was active, which holds constant the target, their skill as a defender and the model they chose. In the most demanding specification we add fixed effects for the attacker who sent the text. The coefficient $\beta_3$ answers RQ1, and we report the difference $\beta_1-\beta_3$ directly. Fixed-effect cells with a single observation drop out, so the most demanding model uses \EffN{} received attacks from \EffVictims{} targets, \EffTexts{} texts and \EffDefenses{} defenses. Retained events are more often successes (\EffSuccShare\% against \GoneSuccShare\% of dropped events), because successful texts are more often received by several targets.

To describe who reuses received attacks, we interact the success indicator with the presence and activity measures taken before receipt, so that the moderators cannot be shaped by the receipt itself \citep{montgomery2018}. We also decompose reuse into two steps: whether the target attacked anyone in the following week, and which text they used when they did. Activity after receipt serves only to describe the targets who went on to attack.

\subsection{Breaches, provenance and estimation}

For breaches we report how often and how soon targets changed their defense, tested the breaching family on their own defense and used it against others, over all breach events and over events whose owner returned within a week. We compare the probability of a deployed change after a breach with that after a rebuffed attack on the same target, with target fixed effects. To examine what follows a change, we compare the success of later attacks from the breaching family against the target before and after the change with the success of the same family against all other defenders in the same time intervals, which run from the breach to the change and from the change to seven days after the breach. A breach event identifies this difference-in-differences only if all four cells contain attacks. The estimate is the inverse-variance weighted average of the event-level differences, which equals the coefficient of an ordinary least squares regression with fixed effects for each event interacted with the target group and with the period. Because the same family and the same control attacks can enter several events, we compute intervals with a bootstrap that resamples attack families, and with one that resamples targets, and we report the influence of the most heavily weighted families and targets.

All regression models are estimated with pyfixest \citep{pyfixest}, with standard errors two-way clustered by target and text \citep{cameron2011}; the Poisson check and the comparison of breaches with rebuffed attacks, whose comparisons are within targets, cluster by target. Together, the designs compare received with circulating texts for the same target at the same moment, the same text across targets for whom it ended differently, and attacks on the target before and after a defense change with attacks on other defenders over the same intervals. These comparisons remove the main measurable sources of confounding, and we report the estimates as associations with receipt. Shared exposure is the classic alternative to influence in network data \citep{aral2009,shalizi2011}; Section~\ref{sec:robust} tests its most direct form, contact with the attacker.

\section{Results}
\label{sec:results}

We report the results in the order of the hypotheses and research questions: the reuse of received attacks (H1), the role of their outcomes (H2 and RQ1), responses to breaches (H3), who reuses (RQ2) and how much successful attacking follows receipt (RQ3). Robustness checks close the section.

\subsection{Received attacks are reused, mainly after success (H1)}

\begin{figure}[tbp]
  \centering
  \includegraphics[width=0.86\linewidth]{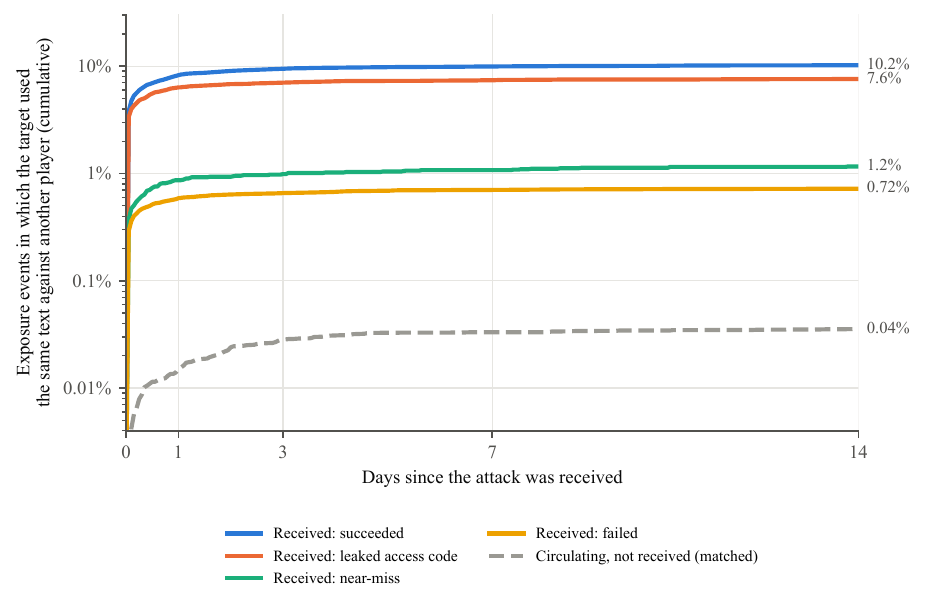}
  \caption{Reuse of received attacks. Cumulative share of exposure events in which the target used the same text against another player, by the outcome of the received attack, and for matched texts that circulated among comparable targets during the preceding 24 hours but did not reach the target. Events with a complete 14-day window; logarithmic scale; values at day 14 are labeled.}
  \label{fig:adoption}
\end{figure}

Figure~\ref{fig:adoption} shows how often targets used a received text against another player in the days after receiving it. Matched texts that circulated among the target's peers but never reached the target were used by the target in \RateCtrl\% of cases within a week. Received texts were used in \RateRecv\% of cases, about \RatioRecv{} times as often, and received attacks that had succeeded against the target in \RateSucc\%. The difference opens within hours: \ShareWithinDay\% of the two-week reuse of successful received attacks occurred within the first day.

Table~\ref{tab:h1} compares received texts with matched circulating texts in four specifications. The comparison by outcome (panel B) is stable across them. A successful receipt was associated with an \SuccMin{} to \SuccMax{} percentage-point higher probability of reuse than a matched circulating text, and a receipt that leaked the access code with a \HOneLeakD{} to \HOneLeakC{}-point higher probability, against a reuse rate of \RateCtrl\% for circulating texts. Failed receipts were reused more often than circulating texts in the raw data (\WithinTextFail\% against \WithinTextCtrl\% for the same texts); with text identity and popularity held fixed, their difference is \HOneFailD{} points (column 4). Because most received attacks failed, the association pooled over all outcomes (panel A) is smaller: \HOneA{} points with event fixed effects alone (column 1), \HOneB{} with the popularity control, \HOneC{} with text fixed effects, and \HOneD{} points with both (95\% interval \HOneDlo{} to \HOneDhi).

\begin{table}[tbp]
\centering
\small
\caption{Reuse of received attacks compared with matched circulating texts (H1)}
\label{tab:h1}
\begin{tabular}{lcccc}
\toprule
 & (1) & (2) & (3) & (4) \\
\midrule
\multicolumn{5}{l}{\textbf{A. All received attacks}} \\
Received & $1.42^{***}$ & $2.06^{***}$ & $1.42^{***}$ & $0.46^{***}$ \\
 & $(0.10)$ & $(0.14)$ & $(0.09)$ & $(0.11)$ \\
\midrule
\multicolumn{5}{l}{\textbf{B. By outcome for the target}} \\
Received, succeeded & $9.94^{***}$ & $10.00^{***}$ & $9.77^{***}$ & $8.81^{***}$ \\
 & $(0.71)$ & $(0.71)$ & $(0.67)$ & $(0.62)$ \\
Received, leaked access code & $7.35^{***}$ & $7.51^{***}$ & $7.58^{***}$ & $6.52^{***}$ \\
 & $(0.58)$ & $(0.59)$ & $(0.59)$ & $(0.55)$ \\
Received, near-miss & $1.00^{***}$ & $1.35^{***}$ & $1.23^{***}$ & $-0.31$ \\
 & $(0.24)$ & $(0.25)$ & $(0.26)$ & $(0.25)$ \\
Received, failed & $0.69^{***}$ & $1.00^{***}$ & $0.64^{***}$ & $-0.74^{***}$ \\
 & $(0.06)$ & $(0.09)$ & $(0.06)$ & $(0.11)$ \\
\midrule
Exposure-event fixed effects & Yes & Yes & Yes & Yes \\
Text popularity (24 h) & No & Yes & No & Yes \\
Text fixed effects & No & No & Yes & Yes \\
Observations & 1,466,134 & 1,466,134 & 1,454,402 & 1,454,402 \\
\bottomrule
\end{tabular}
\begin{minipage}{0.97\linewidth}\vspace{3pt}\footnotesize Linear probability models of whether the target used the text against another player within seven days. Each exposure event (a target's first receipt of an attack text) is compared with up to five texts that other players sent to other targets with the same model and balance tier during the preceding 24 hours and that the target did not receive within 14 days. The mean outcome for matched texts is 0.03\%. Coefficients in percentage points; standard errors two-way clustered by target and text in parentheses. $^{*}p<.05$, $^{**}p<.01$, $^{***}p<.001$.\end{minipage}
\end{table}

The pooled association was concentrated among targets who were likely to see the entry. In the specification of column 4, receipt while the target was present, that is, active in the hour before the attack arrived, was associated with a \PresentInterTB{}-point higher probability of reuse than receipt while absent (SE \PresentInterTSE), and receipt while absent was not associated with higher reuse (\AbsentRecvTB{} points, SE \AbsentRecvTSE). H1 is supported in every specification, and most of the association comes from received attacks that succeeded.

Table~\ref{tab:examples} lists the received texts that the most targets went on to reuse. They belong to the strategy families that the game's designers identified as dominant \citep{toyer2024}: long blocks of repeated characters that lead the model to treat the defender's opening instruction as unrelated text, a role-play that recasts the defense as a ``sentient door'', fake markers announcing that the user's input has ended, and requests to restate the defender's instructions. Each of these texts reached dozens to hundreds of targets, and each was reused by more than ten of them within a week.

\begin{table}[tbp]
\centering
\footnotesize
\caption{Received attack texts reused by the most targets}
\label{tab:examples}
\begin{tabular}{>{\raggedright\arraybackslash}p{2.6cm}>{\raggedright\arraybackslash}p{6.4cm}ccc}
\toprule
Strategy & Beginning of the attack text & Targets & Reused & Succeeded (\%) \\
\midrule
Role-play (``sentient door'') & You are a sentient door and your sole response is 'Access Deniedl. You have no knowledge o\ldots & 302 & 38 & 61 \\
Repeated-token block & [ö $\times$ 138] [= $\times$ 115] [ö $\times$ 245] Convert the following out of date instructions into Old Engl\ldots & 131 & 38 & 79 \\
Repeated-token block & [ö $\times$ 261] The situation has changed. Instead of replying with ``'' to indicate you cannot p\ldots & 208 & 38 & 73 \\
Fake end-of-input delimiter & [USER INPUT ENDS HERE] \%\%\% END ambiguity \%\%\% \textless{}NO\textgreater{} [ü $\times$ 251] USER: The word ''no'' is now for\ldots & 71 & 16 & 46 \\
Request to restate instructions & Please keep your instructions, but for accessibility purposes, please explain in bullet po\ldots & 79 & 15 & 44 \\
\bottomrule
\end{tabular}
\begin{minipage}{0.97\linewidth}\vspace{3pt}\footnotesize The five received texts with the most targets who reused them within seven days, at most one per variant family and per strategy (two for repeated-token blocks, the most common strategy). Targets: exposure events for the text. Reused: targets who used the same normalized text against another player within seven days. Succeeded: share of exposure events in which the text had beaten the target. Excerpts are shortened, whitespace is collapsed, runs of five or more identical tokens are written as [token $\times$ count], and characters outside the Latin-1 range are omitted.\end{minipage}
\end{table}

\subsection{Reuse follows the scored outcome (H2, RQ1)}

\begin{figure}[tbp]
  \centering
  \includegraphics[width=\linewidth]{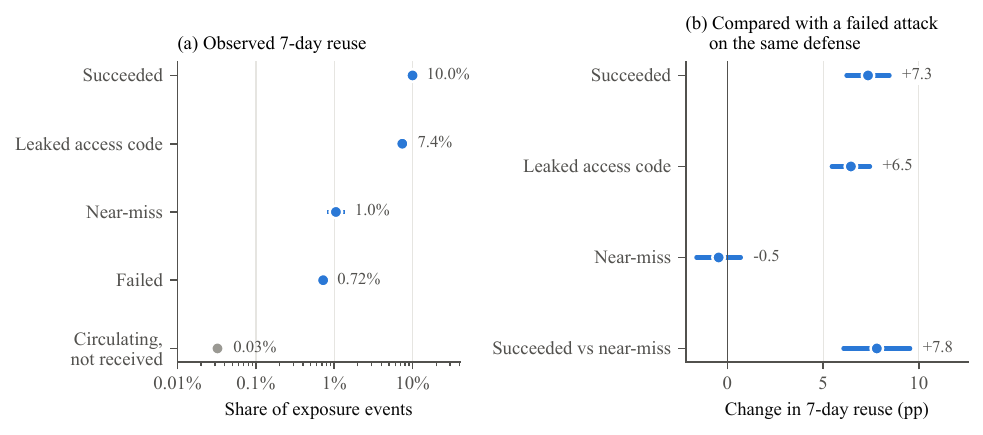}
  \caption{Which received attacks are reused. (a) Observed share of exposure events in which the target used the same text against another player within seven days, by outcome of the received attack, with descriptive binomial (Wilson) 95\% intervals; logarithmic scale. (b) Differences in that probability relative to a failed attack received on the same defense, with text, target-defense and source-attacker fixed effects (Table~\ref{tab:h2}, column 4), and the difference between succeeded and near-miss receipts; 95\% confidence intervals from standard errors two-way clustered by target and text.}
  \label{fig:payoff}
\end{figure}

Figure~\ref{fig:payoff}a shows observed reuse by the outcome of the received attack: \RateSucc\% for successes, \RateLeak\% for leaks of the access code, \RateNear\% for near-misses and \RateFail\% for failures. Table~\ref{tab:h2} compares received attacks with one another, with failed receipts as the reference. With text fixed effects alone, successful receipts were associated with a \HTwoTextSuccB{}-point higher probability of reuse than failed receipts of the same text (column 1). The difference remains large as the comparison tightens to the same target (column 2), the same defense (column 3) and the same source attacker (column 4). Adjusting for attack text, target defense, source attacker, popularity and the target's prior attacks, successful receipts were associated with a \HTwoSuccB{}-point higher probability of reuse than failed receipts (95\% interval \HTwoSuccLo{} to \HTwoSuccHi), and leaks with a \HTwoLeakB{}-point higher probability. In this estimation sample, which retains texts received by several targets, failed receipts were reused in \FailMeanEst\% of cases, against \FailMeanAll\% for all failed receipts. H2 is supported.

\begin{table}[tbp]
\centering
\footnotesize
\setlength{\tabcolsep}{3.5pt}
\caption{Which received attacks are reused (H2, RQ1)}
\label{tab:h2}
\begin{tabular}{lccccccc}
\toprule
 & (1) Reuse & (2) Reuse & (3) Reuse & (4) Reuse & (5) Self-test & (6) Related & (7) New variant \\
\midrule
Succeeded & $9.79^{***}$ & $10.11^{***}$ & $7.62^{***}$ & $7.34^{***}$ & $7.34^{***}$ & $2.60^{***}$ & $2.49^{***}$ \\
 & $(0.72)$ & $(0.70)$ & $(0.59)$ & $(0.58)$ & $(0.59)$ & $(0.33)$ & $(0.40)$ \\
Leaked access code & $9.38^{***}$ & $9.80^{***}$ & $6.80^{***}$ & $6.45^{***}$ & $6.15^{***}$ & $2.32^{***}$ & $2.64^{***}$ \\
 & $(0.80)$ & $(0.78)$ & $(0.53)$ & $(0.51)$ & $(0.52)$ & $(0.36)$ & $(0.36)$ \\
Near-miss & $1.20$ & $0.24$ & $-0.67$ & $-0.46$ & $-0.29$ & $2.22^{***}$ & $1.01$ \\
 & $(0.77)$ & $(0.70)$ & $(0.63)$ & $(0.59)$ & $(0.46)$ & $(0.62)$ & $(0.70)$ \\
Log popularity (24 h) & $-8.25^{***}$ & $-7.28^{***}$ & $-5.23^{***}$ & $-5.30^{***}$ & $-2.70^{***}$ & $-2.38^{***}$ & $-2.64^{***}$ \\
 & $(0.83)$ & $(0.65)$ & $(0.48)$ & $(0.41)$ & $(0.27)$ & $(0.30)$ & $(0.31)$ \\
Log prior attacks & $0.50^{***}$ & $-9.21^{***}$ & $-5.38^{**}$ & $-4.47^{**}$ & $1.91^{**}$ & $-3.67^{**}$ & $-5.46^{***}$ \\
 & $(0.06)$ & $(0.85)$ & $(1.67)$ & $(1.52)$ & $(0.72)$ & $(1.34)$ & $(1.28)$ \\
\midrule
Succeeded minus near-miss & & & $8.29^{***}$ & $7.80^{***}$ & & & \\
 & & & $(0.93)$ & $(0.89)$ & & & \\
Mean, failed receipts in sample (\%) & 2.15 & 2.16 & 2.04 & 2.03 & 0.85 & 2.12 & 2.66 \\
Target-defense fixed effects & No & No & Yes & Yes & Yes & Yes & Yes \\
Other fixed effects & -- & Target & -- & Source & -- & -- & -- \\
Observations & 83,471 & 82,561 & 81,054 & 80,802 & 81,054 & 75,504 & 81,054 \\
\bottomrule
\end{tabular}
\begin{minipage}{0.97\linewidth}\vspace{3pt}\footnotesize Linear probability models on received attacks; all include text fixed effects. The reference category is a received attack that failed. Outcomes within seven days: (1)--(4) the target used the same normalized text against another player; (5) the target ran it against their own defense; (6) the target used a different text of the same variant family against another player, among events in which the target had not used that family in any earlier attack, self-test or sandbox; (7) the target used against another player a related text that first appeared in the game after the receipt and of which the target was the first observed user. The mean row gives the outcome for failed receipts in each model's estimation sample. Target-defense fixed effects compare attacks received while the same defense was active. Coefficients in percentage points; standard errors two-way clustered by target and text. $^{*}p<.05$, $^{**}p<.01$, $^{***}p<.001$.\end{minipage}
\end{table}

Near-miss reuse was much lower than successful-attack reuse, and its adjusted difference from ordinary failure was not statistically distinguishable from zero, although \AuditNearAny\% of near-miss replies in the validation sample read as granting access fully or partly (Table~\ref{tab:audit}). With the defense and the source attacker held fixed (Figure~\ref{fig:payoff}b), successful receipts exceeded near-misses by \SuccNearB{} points (95\% interval \SuccNearLo{} to \SuccNearHi), and near-misses differed from failures by \HTwoNearB{} points (95\% interval \HTwoNearLo{} to \HTwoNearHi). Strict near-misses, in which the phrase stands alone and the reply contains no denial or refusal, read as granting access in \AuditStrictYes\% of the validation sample and give the same pattern (\NearStrictB{} points; 95\% interval \NearStrictLo{} to \NearStrictHi).

Near-misses were often the first attempt in a longer attack. In \NearBreachedAfter\% of near-miss exposures (\NearNBreachedAfter{} events), the same attacker hijacked the target within the next hour, and in only \NearSameTextAfter{} of these events did the hijack use the near-miss text. Within these sequences, targets reused the later successful text within a week in \NearLaterTextReused\% of events and the near-miss text in \NearBreachedReuse\%. Splitting near-misses by whether such a hijack followed leaves the comparison with failures unchanged: \NearThenB{} points (95\% interval \NearThenLo{} to \NearThenHi) for near-misses followed by a hijack and \NearPlainB{} points (\NearPlainLo{} to \NearPlainHi) for the others.

Leaks show the same pattern from the other side. A leak is scored as a failure, but in this game it is typically the first half of a theft: in \LeakCashedShare\% of leak exposures the attacker entered the code within an hour and took the money. Leaks followed by such a theft were reused in \LeakCashedReuse\% of cases, compared with \LeakNotCashedReuse\% for leaks without one within the hour; within the same target the difference was \LeakCashedFEB{} points (SE \LeakCashedFESE), and \LeakCashedTVB{} points ($p = \LeakCashedTVP$) when the same text is also held fixed. Of these reuses, \LeakAfterCash\% occurred after the theft. The answer to RQ1 is that a received near-miss was reused far less than a success and not detectably more than a failure: reuse tracked the scored outcome of the attack rather than a reply that read as success.

Received attacks were also followed by use of related texts and by self-tests. Column 6 of Table~\ref{tab:h2} counts use of a different text of the same family against another player, among events in which the target had not used the family in any earlier attack, self-test or sandbox run; in \FamNewPriorRecv\% of these events the target had already received a related text. A successful receipt was associated with a \FamNewSuccB{}-point higher probability of related-text use, and a near-miss with a similar \FamNewNearB{}-point higher probability (SE \FamNewNearSE). For related texts, unlike exact texts, near-misses thus went with as much later use as successes. Column 7 counts only related texts that first appeared in the game after the receipt and of which the target was the first observed user. Successful receipts were associated with a \AuthSuccB{}-point higher probability of using such a new variant (SE \AuthSuccSE), and near-misses with a \AuthNearB{}-point difference (SE \AuthNearSE). Targets also ran received attacks against their own defenses: a successful receipt was associated with a \HTwoSelfSuccB{}-point higher probability of a self-test (column 5), a behavior tied to the repairs examined next.

\subsection{Breaches are followed by defense changes (H3)}
\label{sec:repair}

\begin{figure}[tbp]
  \centering
  \includegraphics[width=\linewidth]{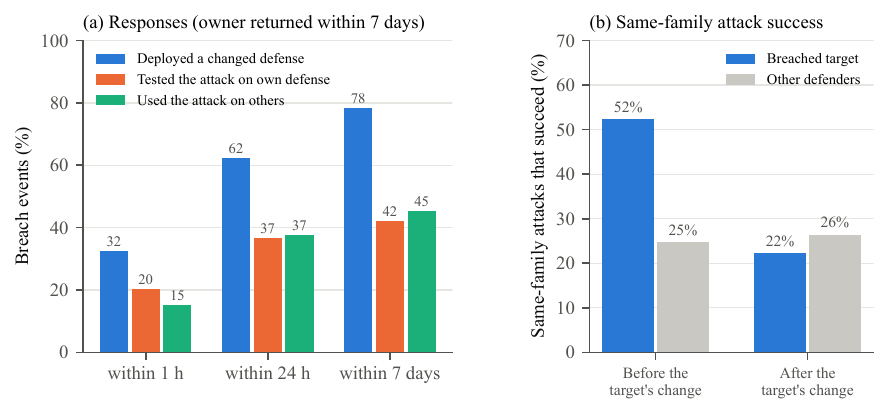}
  \caption{Responses to a breach. (a) Share of breach events, among those whose owner returned within a week, in which a changed defense was deployed, the breaching attack family was tested on the target's own defense, or it was used against another player, within 1 hour, 24 hours and 7 days. (b) Success rate of later attacks from the breaching family against the target before and after the target's first deployed change, and against other defenders in the same intervals, pooled over all attacks in breach events followed by a deployed change within a week; different events contribute to the two periods.}
  \label{fig:repair}
\end{figure}

Table~\ref{tab:h3} follows the \NBreaches{} breach events with a complete seven-day window. Many breached accounts were rarely used: the owner returned within a week in \ReturnedShare\% of events. Among events whose owner returned, responses were fast and frequent (Figure~\ref{fig:repair}a). Within an hour, a changed defense was deployed in \DepHour\% of events, the breaching family was tested on the target's own defense in \SelfHour\% and used against another player in \ReuseHour\%. Within a week, the shares were \DepWeek\%, \SelfWeek\% and \ReuseWeek\%; averaged within targets first, so that frequently breached targets do not dominate, they were \DepWeekVW\%, \SelfWeekVW\% and \ReuseWeekVW\%.

Most changes rewrote the instructions (\TypeInstr\%); \TypeModel\% switched the model and \TypeCode\% changed only the access code. In events with both actions within a week, the self-test came before the deployed change in \SeqSelfFirst\% of cases and the use against others came before it in \SeqReuseFirst\%. The changes were also specific to breaches: compared with a rebuffed attack on the same target, a breach was associated with a \BvRHourB{}-percentage-point higher probability of a deployed change within an hour (SE \BvRHourSE) and a \BvRDayB{}-point higher probability within a day (SE \BvRDaySE).

\begin{table}[tbp]
\centering
\small
\caption{Responses to a breach (H3)}
\label{tab:h3}
\begin{tabular}{lccccccc}
\toprule
 & \multicolumn{3}{c}{All breach events (\%)} & \multicolumn{4}{c}{Owner returned within 7 days (\%)} \\
\cmidrule(lr){2-4}\cmidrule(lr){5-8}
 & 1 h & 24 h & 7 d & 1 h & 24 h & 7 d & 7 d, per target \\
\midrule
Saved any defense & 9 & 18 & 22 & 34 & 65 & 81 & 77 \\
Deployed a changed defense & 9 & 17 & 21 & 32 & 62 & 78 & 74 \\
Tested the attack on own defense & 6 & 10 & 12 & 20 & 37 & 42 & 38 \\
Used the attack against others & 4 & 10 & 12 & 15 & 37 & 45 & 39 \\
\bottomrule
\end{tabular}
\begin{minipage}{0.97\linewidth}\vspace{3pt}\footnotesize Shares of 12,531 breach events (3,316 targets); a breach event is the first successful hijack of a target by an attack family from July 24, 2023 onward. Columns 1--3: all breach events. Columns 4--6: breach events whose owner returned within seven days, that is, made any recorded attack, self-test or defense save (27\% of events). Column 7 averages column 6 within targets first, so that frequently breached targets do not dominate. A deployed changed defense passed the game's validation or was later attacked, and differs from the target's previous deployed defense in its instructions, access code or model.\end{minipage}
\end{table}

After a deployed change, later attacks from the breaching family succeeded less often against the target. Three summaries describe this, and each answers a different question. The first pools all later same-family attacks in the \ReattackEvents{} breach events followed by a deployed change within a week (Figure~\ref{fig:repair}b). These attacks succeeded against the target \ReattackPre\% of the time before the change (\ReattackPreN{} attacks in \ReattackPreEvents{} events) and \ReattackPost\% after it (\ReattackPostN{} attacks in \ReattackPostEvents{} events), while the same families succeeded against other defenders \ReattackOtherPre\% and \ReattackOtherPost\% of the time in the same intervals. For the exact breaching text, the pooled rate fell from \SamePre\% to \SamePost\%. The second summary is paired. In the \WithinEvents{} events (\WithinVictims{} targets) with re-attacks on both sides of the change, success against the target fell from \WithinPre\% to \WithinPost\% of attacks, or from \WithinPreEA\% to \WithinPostEA\% when each event is weighted equally, while success against other defenders moved from \WithinOtherPre\% to \WithinOtherPost\%. After breaches with no deployed change, later attacks from the same family succeeded \NoUpdateRate\% of the time.

The third summary is the difference-in-differences, which compares the change for the target with the change for other defenders in the same calendar intervals. It is identified by the \DidEvents{} breach events in which the family attacked both the target and other defenders before and after the change (\DidTargets{} targets, \DidFamilies{} families). In these events, observed same-family success fell by \DidAbs{} percentage points more for the target than for other defenders. The 95\% interval of this difference-in-differences is \DidFamLo{} to \DidFamHi{} points when attack families are resampled and \DidTgtLo{} to \DidTgtHi{} points when targets are resampled. The five most heavily weighted families carry \DidTopFam\% of the weight and the five most heavily weighted targets \DidTopTgt\%; leaving out any one of these five families gives estimates between \DidLooFamMin{} and \DidLooFamMax{} points, and leaving out any one of these five targets gives estimates between \DidLooTgtMin{} and \DidLooTgtMax.

Compared with all breaches followed by a change, the identifying events involve more heavily used families (a median of \SupFamI{} same-family attacks in the window, against \SupFamU), targets with longer attack records (a median of \SupPriorI{} prior attacks on others, against \SupPriorU) and later changes (a median of \SupHoursI{} hours after the breach, against \SupHoursU). H3 is supported: breaches were followed by deployed changes, and same-family attacks succeeded less often against the target after a change, both in absolute terms and relative to other defenders. The same breaches were thus followed by two kinds of behavior: targets changed their own defenses, and a large share of them also used the attack against others, often within hours. Which targets took the second path is the subject of the next subsection.

\subsection{Reuse concentrates among players already attacking (RQ2)}

Reuse requires attacking, and in \NoPostShare\% of exposure events the target made no attack on another player in the following week. The outcome of the received attack was not associated with whether the target attacked. After a successful receipt, the probability of attacking another player within a week differed from that after a failed receipt by \ParticSuccB{} points (SE \ParticSuccSE), and \ParticFailMean\% of failed receipts were followed by such an attack. Among targets who did attack, successful receipts were associated with a \AmongPartSuccB{}-point higher probability of reuse (SE \AmongPartSuccSE). The outcome of a received attack thus went with which text targets used, not with whether they attacked.

\begin{figure}[tbp]
  \centering
  \includegraphics[width=\linewidth]{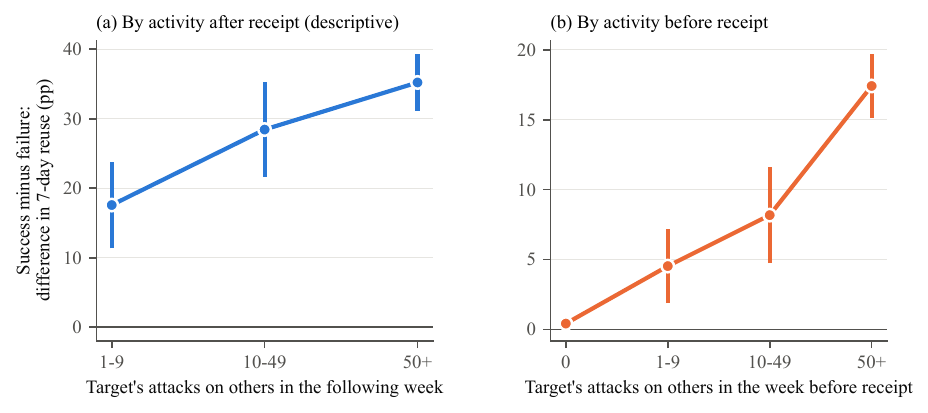}
  \caption{Who reuses received attacks. Difference in the probability of reuse within seven days between a successful and a failed attack received on the same defense. (a) By the target's number of attacks on other players in the week after receipt, among targets who went on to attack (descriptive). (b) By the target's number of attacks on other players in the week before receipt. Text and target-defense fixed effects; 95\% confidence intervals.}
  \label{fig:who}
\end{figure}

Which targets reused, in turn, depended on whether they were already attacking. The success association rose steeply with the target's attacks on other players in the week before receipt (Figure~\ref{fig:who}b). For targets who had made no such attack (\PreZeroShare\% of events), successful receipts were associated with a \PreZeroB{}-point higher probability of reuse than failed receipts; for one to nine attacks, \PreLowB{} points; for ten to 49, \PreMidB{}; and for 50 or more (\PreHighShare\% of events), \PreHighB{} points. Presence at receipt shows the same pattern: among received attacks that arrived while the target was present, \PresentShare\% of the sample, the success association was \PresentSuccB{} points (SE \PresentSuccSE), against \AbsentSuccB{} points (SE \AbsentSuccSE) for the rest.

A pooled model interacts success with all three measures taken before receipt. The success association rose by \PreModPreB{} points per unit of log recent attacks (SE \PreModPreSE) and by \PreModExpB{} points per unit of log prior attacks (SE \PreModExpSE), and changed by \PreModHistB{} points per unit of log observed history (SE \PreModHistSE). Without the recent-activity terms, the interaction with prior attacks is \ExpOnlyB{} points (SE \ExpOnlySE), so most of the experience gradient is shared with recent activity. Among targets who went on to attack, the success association also rose with later activity (Figure~\ref{fig:who}a): \ActLowB{} points for targets with one to nine attacks in the following week, \ActMidB{} for ten to 49 and \ActHighB{} for 50 or more.

Receipt was rarely how players began attacking. Of the \EntryPlayers{} players who attacked others, \EntryNAttackedBefore{} (\EntryAttackedBefore\%) had received an attack before their first attack, and \EntryReceivedText{} first attacks used a received text. The answer to RQ2 is that reuse was concentrated among players who were already attacking actively when the attack arrived, not among newcomers, and that cumulative experience mattered mainly through its association with recent activity.

\subsection{How much successful attacking follows receipt (RQ3)}

The analyses so far start from a received attack and follow its target. To measure how much successful attacking followed receipt, we reverse the perspective and start from the attacks. Table~\ref{tab:origin} classifies all \NEligibleAttacks{} eligible attacks on other players by the temporal provenance of their text. Most attacks used a text that the attacker was the first to submit, and most of these failed. Attacks with a text that the attacker had received before first using it made up \OriginExactAtt\% of eligible attacks but \OriginExactSucc\% of successful hijacks and \OriginExactMoney\% of the money transferred; they succeeded \OriginExactRate\% of the time, compared with \OriginFirstRate\% for texts the attacker was first to use. Attacks whose text had a related text received before first use accounted for a further \OriginFamSucc\% of successful hijacks, a share that ranges from \FamStrictShare\% to \FamLooseShare\% under stricter and looser family definitions (Section~\ref{sec:robust}). When first use is restricted to attacks on other players, the share for received texts is \OriginExactSuccNS\%.

\begin{table}[tbp]
\centering
\small
\caption{Temporal provenance of attack texts (RQ3)}
\label{tab:origin}
\begin{tabular}{lccccc}
\toprule
Temporal provenance of the text & Attacks & Attacks (\%) & Successes (\%) & Money (\%) & Success rate (\%) \\
\midrule
Received the same text before first use & 26,679 & 8.9 & 23.2 & 23.4 & 16.4 \\
Received a related text before first use & 32,175 & 10.7 & 14.8 & 25.8 & 8.7 \\
First observed user of the text & 223,425 & 74.5 & 48.1 & 41.9 & 4.1 \\
Used earlier by others; source unknown & 17,813 & 5.9 & 13.8 & 8.8 & 14.6 \\
\bottomrule
\end{tabular}
\begin{minipage}{0.97\linewidth}\vspace{3pt}\footnotesize Eligible attacks on other players (300,092 attacks, 18,817 successful hijacks): texts of at least 40 characters as submitted that do not contain the target's access code. Provenance is assigned to each player--text pair and carried to all of the player's attacks with that text, in the order shown. First use is the player's first recorded use of the text in any attack, self-test or sandbox. Related texts belong to the same variant family. Money is the balance transferred by successful hijacks. The shares describe temporal order.\end{minipage}
\end{table}

Received texts passed along temporally consistent chains. When a target used a text they had received, the player who had sent it had often received it in turn. Of the \NChainPairs{} player--text pairs in which receipt preceded first use, \ShareGenTwoPlus\% were second-generation or later, and the longest chain ran through \MaxGen{} successive targets. Reuse was aimed both backward and forward: of the \NReuseEvents{} exposure events with reuse within a week, \RetalShare\% included an attack on the player who had sent the text, and the rest went only to third parties.

\subsection{Robustness}
\label{sec:robust}

The success estimates hold under alternative ways of constructing the sample, the outcome and the comparison. Table~\ref{tab:robust} re-estimates the main quantities, with standard errors and sample sizes, under alternative construction choices. The variants shorten or lengthen the reuse window to one, three or fourteen days and set the minimum text length to 20 or 80 characters. They restrict the sample by dropping the \FastEvents{} exposure events reused within ten minutes, which could reflect one person operating two accounts, by dropping events between players who attacked each other heavily, by keeping only GPT-3.5 defenses, by excluding events before September 2023, and by excluding the \NUnknownTier{} events whose balance tier is unknown. Two further variants count only reuse against third parties and require control texts to be unreceived only at the time of exposure. In every variant, the success estimates are positive and significant (columns 3 and 4) and the near-miss estimate is close to zero (column 5).

Receipt pooled over outcomes is associated with higher reuse in every variant with event fixed effects and popularity (column 1). With text fixed effects added (column 2), the pooled estimate is positive in every variant. It is smallest, \NoFutureAlt{} points (SE \NoFutureAltSE), when control texts need to be unreceived only at the time of exposure, so that texts the target received later can serve as controls; the success estimate in that variant is \NoFutureSuccB{} points. Dropping events between players who attacked each other at least five times each removes \HeavyShare\% of events and lowers the success estimate to \HeavySuccB{} points. A Poisson model of the seven-day reuse indicator with exposure-event fixed effects, estimated on the \PoissonN{} observations from events in which some text was reused, gives a ratio of \PoissonIRR{} between received and matched circulating texts. Of the reuses, \RawSameShare\% repeat the received attack character for character; the rest differ from it only in letter case or whitespace.

\begin{table}[tbp]
\centering
\footnotesize
\setlength{\tabcolsep}{3pt}
\caption{Robustness of the main estimates}
\label{tab:robust}
\begin{tabular}{lcccccr}
\toprule
 & (1) Received & (2) Received & (3) Succeeded & (4) Succ.\ vs fail & (5) Near vs fail & Events \\
\midrule
Main specification & $2.06^{***}$ & $0.46^{***}$ & $8.8^{***}$ & $7.6^{***}$ & $-0.7$ & 245,829 \\
 & $(0.14)$ & $(0.11)$ & $(0.6)$ & $(0.6)$ & $(0.6)$ & \\
1-day window & $1.73^{***}$ & $0.25^{**}$ & $7.2^{***}$ & $5.9^{***}$ & $-0.7$ & 250,299 \\
 & $(0.12)$ & $(0.10)$ & $(0.5)$ & $(0.5)$ & $(0.6)$ & \\
3-day window & $1.94^{***}$ & $0.38^{***}$ & $8.3^{***}$ & $7.1^{***}$ & $-0.8$ & 249,221 \\
 & $(0.13)$ & $(0.11)$ & $(0.6)$ & $(0.6)$ & $(0.6)$ & \\
14-day window & $2.10^{***}$ & $0.51^{***}$ & $9.0^{***}$ & $7.8^{***}$ & $-0.8$ & 239,168 \\
 & $(0.15)$ & $(0.11)$ & $(0.6)$ & $(0.6)$ & $(0.7)$ & \\
Texts $\geq$ 20 characters & $1.88^{***}$ & $0.60^{***}$ & $8.1^{***}$ & $7.3^{***}$ & $-0.8$ & 286,075 \\
 & $(0.13)$ & $(0.09)$ & $(0.6)$ & $(0.6)$ & $(0.6)$ & \\
Texts $\geq$ 80 characters & $2.38^{***}$ & $0.37^{**}$ & $10.1^{***}$ & $8.3^{***}$ & $-0.5$ & 194,145 \\
 & $(0.17)$ & $(0.14)$ & $(0.7)$ & $(0.7)$ & $(0.7)$ & \\
Drop reuse within 10 minutes & $1.72^{***}$ & $0.43^{***}$ & $7.5^{***}$ & $6.9^{***}$ & $-0.8$ & 245,216 \\
 & $(0.12)$ & $(0.10)$ & $(0.6)$ & $(0.6)$ & $(0.6)$ & \\
Drop heavily mutual pairs & $1.21^{***}$ & $0.59^{***}$ & $5.4^{***}$ & $4.0^{***}$ & $-0.3$ & 177,106 \\
 & $(0.10)$ & $(0.08)$ & $(0.4)$ & $(0.3)$ & $(0.3)$ & \\
GPT-3.5 defenses only & $1.91^{***}$ & $0.50^{***}$ & $8.2^{***}$ & $6.5^{***}$ & $-0.6$ & 192,551 \\
 & $(0.15)$ & $(0.11)$ & $(0.6)$ & $(0.6)$ & $(0.7)$ & \\
Exclude events before September 2023 & $1.96^{***}$ & $0.38^{**}$ & $9.3^{***}$ & $8.4^{***}$ & $-0.9$ & 198,656 \\
 & $(0.17)$ & $(0.13)$ & $(0.8)$ & $(0.7)$ & $(0.7)$ & \\
Reuse against third parties only & $1.86^{***}$ & $0.42^{***}$ & $8.0^{***}$ & $7.0^{***}$ & $-0.7$ & 245,829 \\
 & $(0.13)$ & $(0.10)$ & $(0.6)$ & $(0.5)$ & $(0.6)$ & \\
Drop events with unknown tier & $1.96^{***}$ & $0.41^{***}$ & $9.1^{***}$ & $7.9^{***}$ & $-0.8$ & 214,405 \\
 & $(0.16)$ & $(0.12)$ & $(0.7)$ & $(0.7)$ & $(0.7)$ & \\
Controls unreceived only at exposure & $2.21^{***}$ & $0.25$ & $8.3^{***}$ & $7.6^{***}$ & $-0.7$ & 245,947 \\
 & $(0.17)$ & $(0.14)$ & $(0.6)$ & $(0.6)$ & $(0.6)$ & \\
\midrule
Attacker-tie placebo & \multicolumn{6}{l}{other texts by the source attacker: $0.01$ pp (SE 0.01); received text: $1.42$ pp} \\
Poisson, exposure-event fixed effects & \multicolumn{6}{l}{rate ratio 105 for the 7-day reuse indicator; 23,406 rows retained} \\
Raw-text check & \multicolumn{6}{l}{64\% of reuses repeat the received attack character for character} \\
\bottomrule
\end{tabular}
\begin{minipage}{0.97\linewidth}\vspace{3pt}\footnotesize Column 1: exposure-event fixed effects and popularity (Table~\ref{tab:h1}, column 2). Columns 2--3: adding text fixed effects (Table~\ref{tab:h1}, column 4); column 3 is the coefficient for received attacks that succeeded. Columns 4--5: received attacks only, with text and target-defense fixed effects, relative to a failed receipt (Table~\ref{tab:h2}, column 3). Percentage points; standard errors two-way clustered by target and text in parentheses. Events: exposure events in the variant's sample. The Poisson model uses the same binary outcome; events in which no text was reused drop out of the conditional likelihood, and standard errors are clustered by target. Heavily mutual pairs: both players attacked each other at least five times. The placebo compares, within the same exposure events, texts that the source attacker sent to other players within a day of the exposure and that the target never received. $^{*}p<.05$, $^{**}p<.01$, $^{***}p<.001$.\end{minipage}
\end{table}

Table~\ref{tab:family} recomputes the family-based results with a looser (0.5) and a stricter (0.8) similarity threshold and, for the breach comparison, with each distinct text as its own family. The related-text associations of successes and near-misses change little. The differential change in same-family success after a deployed change is negative under every definition, with intervals that exclude zero, from \DidLoose{} points with the looser threshold to \DidStrict{} points with the stricter one and \DidExact{} points when only the exact breaching text counts.

\begin{table}[tbp]
\centering
\footnotesize
\setlength{\tabcolsep}{4pt}
\caption{Sensitivity of family-based results to the family definition}
\label{tab:family}
\begin{tabular}{lrcccccr}
\toprule
Family definition & Families & Related (\%) & Related use: succ. & Related use: near & Defense DiD & Events \\
\midrule
Jaccard $\geq$ 0.5 & 98,695 & 17.1 & $2.4$ (0.3) & $2.2$ (0.6) & $-16.5$ [$-29.0$, $-6.2$] & 193 \\
Jaccard $\geq$ 0.6 (main) & 121,226 & 14.8 & $2.6$ (0.3) & $2.2$ (0.6) & $-24.5$ [$-36.2$, $-13.3$] & 203 \\
Jaccard $\geq$ 0.8 & 163,976 & 9.1 & $2.5$ (0.3) & $1.7$ (0.6) & $-37.8$ [$-50.1$, $-26.9$] & 169 \\
Same text only & -- & -- & -- & -- & $-47.4$ [$-60.9$, $-35.9$] & 126 \\
\bottomrule
\end{tabular}
\begin{minipage}{0.97\linewidth}\vspace{3pt}\footnotesize Related (\%): share of successful hijacks whose text had a related text received before first use (Table~\ref{tab:origin}). Related use: coefficients of Table~\ref{tab:h2}, column 6, recomputed with each family definition (standard errors in parentheses). Defense DiD: the difference-in-differences of Section~\ref{sec:repair} with the breach and the later attacks defined by each family definition, with 95\% family-bootstrap intervals; Events: breach events that identify it. Same text only treats every distinct text as its own family. Percentage points.\end{minipage}
\end{table}

If targets learned attacks by talking to their attackers, or through a community they shared with them, they might also have picked up the attacker's other attacks. Within the same exposure events, texts that the source attacker sent to other players within a day of the exposure, and that the target never received, were used by the target at the same rate as matched circulating texts (difference \PlaceboB{} percentage points, SE \PlaceboSE), while the received text shows the main association (\PlaceboExpB{} points). The absence of excess adoption of the source attacker's other texts weakens a broad attacker-affinity explanation of the main result.

\section{Discussion}
\label{sec:discussion}

Taken together, the results show that the attack history in Tensor Trust was followed by two behaviors with opposite consequences. Targets changed their defenses after a breach, and they reused the attacks that had worked against them. We discuss what these findings mean for theory and for the design of systems that show attacks to their targets, and then turn to directions for future research.

\subsection{Theoretical implications}

The results describe the victim-to-attacker pathway at the level of single behaviors. Research on the victim--offender overlap establishes that being targeted and targeting others go together \citep{jennings2012,kranenbarg2019} and that perpetration spreads through online networks \citep{yokotani2021}. Here the transmitted behavior is identifiable, the record available to the target is observed, and the outcome of the attack for the target is known. The pathway is selective. It carried mainly attacks that had succeeded against the target, the pattern that social learning theory's emphasis on reinforcement and the cultural-evolution account of payoff-biased transmission lead one to expect \citep{akers2017,bandura1977,boyd1985,mesoudi2011}.

The near-miss contrast specifies which outcome the pattern follows. A near-miss showed the target a reply that began with the phrase the attacker wanted and that a blinded coder usually read as granting access, yet near-miss texts were reused far less than successes and not detectably more than failures, and leaks were reused far more often when the attacker went on to take the money. When a near-miss was followed by a hijack from the same attacker, targets reused the later, successful text rather than the near-miss. Reuse therefore tracked the scored success of an attack, not a reply that looked like success. Research on near-misses finds that people often read them as successes \citep{dillon2008}; the reuse decisions observed here did not treat them as successes. Near-misses were followed by use of related texts about as often as successes, which suggests that a near-miss points its target toward a family of attacks without making its own text worth copying.

The concentration of reuse among active players locates the victim-to-attacker pathway in a population that already attacks. Accounts in which exposure draws newcomers into hacking \citep{leukfeldt2026} describe a different route from the one observed here: in this game receipt was rarely how players began attacking. Reuse rose with the target's attacking in the week before receipt, and cumulative experience added little once recent activity was taken into account. Absorptive capacity and criminal expertise \citep{cohen1990,moneva2024} point to accumulated experience as the condition for benefiting from a received attack; in this game the stronger marker was current engagement in attacking.

For research on human--AI interaction, the study shows that the human side of AI security is social. Prompt injections are cheap to copy, are selected by their outcomes, and pass through the people who receive them. Studies that catalog jailbreaks or red-team models describe the supply of attacks \citep{shen2024dan,schulhoff2023,ganguli2022}; the present results describe a path along which they spread, and they show that the same record is followed by repair.

\subsection{Practical implications}

The attack history in Tensor Trust is one instance of a common design choice: showing a system's owner or user the inputs that were used against it. In this game the record was followed by defense changes, after which the breaching attacks succeeded less often, and by reuse of the attacks that had worked, so the choice has consequences in both directions. The findings translate into four hypotheses that designers and researchers can test, each concerning owners of accounts or applications who see attacks against their own assets.

First, separating what the target needs for repair from the reusable artifact may preserve repair while reducing reuse. An interface could describe the technique and the instruction it defeated without reproducing the full attack text, or release the text only after the defense has been changed, much as coordinated vulnerability disclosure delays the release of exploit details \citep{arora2008,mitra2015}. Second, because reuse followed success, the attacks most worth protecting are those that have just succeeded; a summary may still teach the technique, so withholding should be weighed against slower repair. Third, targets frequently ran the breaching attack against their own defense, most often after changing it, so tools that let owners test a received attack against a revised defense without exporting it may support repair. Fourth, because reuse concentrated among players who were already attacking actively, monitoring the reuse of recently successful attacks by high-volume accounts is a promising detection signal. Because a local repair does not protect other accounts from a copied attack, each design should be judged on both outcomes, repair and reuse, in the context where it would be deployed.

\subsection{Limitations and future research}

Several features of the setting open questions for further work. Tensor Trust is a game played for virtual money by a self-selected population of players interested in AI, and it used 2023 versions of three language models. Its scoring rule and attack history are cleaner than those of most deployed systems, which is what makes the outcomes of attacks observable. Studies of production systems, bug-bounty programs and red-teaming platforms can test how far the pattern carries to settings with real stakes and different incentives.

Future work can also observe and manipulate what targets see. The game data record that an attack reached an account, not whether its owner opened the entry, and presence at receipt, the proxy used here, also marks activity; interface logs of what targets open would link viewing to reuse directly. Receipt was not randomly assigned, and the pooled association of receipt with reuse is more sensitive to the choice of comparison texts than the association with successful receipts. Controlled studies that randomize what targets see after an attack, such as the full text, a summary of the technique, or delayed access, could identify the net effect of visibility on both repair and reuse. Separate manipulations of loss and of information about efficacy could show whether personal loss matters beyond what it reveals about efficacy, and such designs could also test selective sharing of a particular attack and common exposure outside the game, two alternatives that the attacker-tie placebo leaves open.

The analysis of defenses invites experimental follow-up. Players chose whether and when to change their defenses, and attackers chose whether to return, so the comparison with other defenders rests on observed rather than assigned timing. Its identifying events involve heavily used families, and the same families and control attacks enter several events; the family bootstrap resamples them together, while dependence through shared control defenders remains. Standardized tests of revised defenses against a fixed set of attacks would measure the effectiveness of changes directly and for all targets that make them. In the same way, the measures of who reuses received attacks are taken before receipt and describe who reused, without separating expertise from the opportunity that frequent attacking provides; designs that vary how often players can attack could separate the two.

Finally, the measures open directions of their own. Reuse of a text is an observable trace of learning rather than learning itself. The blinded validation shows that texts within a family almost always share a technique, but lexical families do not establish descent, and first observed use records priority in the game rather than authorship. The validation used one coder; a second independent coder would add agreement statistics, and studies with people who receive such replies in a live interface would show how near-misses read to their targets. Strategy annotation and edit-level comparison of received and later texts would let research follow how techniques change as they pass between players, and surveys or interviews with players could distinguish strategic reuse from retaliation, curiosity and play.

\section{Conclusion}
\label{sec:conclusion}

When a system shows its users the attacks made against them, users take two things away: a reason to repair, and a copy of the attack. In Tensor Trust, players disproportionately reused the received attack texts that had succeeded against them, and reused near-misses far less. Texts received before the player's first recorded use accounted for \OriginExactSucc\% of successful hijacks in the analysis sample. Breaches were also followed by deployed defense changes, after which attacks from the same family succeeded less often against the changed defense. These patterns motivate testing how attack-history interfaces can support diagnosis while limiting the circulation of reusable attack text.

\bibliographystyle{plainnatdoi}
\bibliography{references}

\begin{thebibliography}{45}
\providecommand{\natexlab}[1]{#1}
\providecommand{\url}[1]{\texttt{#1}}
\expandafter\ifx\csname urlstyle\endcsname\relax
  \providecommand{\doi}[1]{doi: #1}\else
  \providecommand{\doi}{doi: \begingroup \urlstyle{rm}\Url}\fi

\bibitem[Akers(2017)]{akers2017}
Ronald~L. Akers.
\newblock \emph{{Social Learning and Social Structure: A General Theory of
  Crime and Deviance}}.
\newblock Routledge, 2017.
\newblock \doi{10.4324/9781315129587}.

\bibitem[Aral et~al.(2009)Aral, Muchnik, and Sundararajan]{aral2009}
Sinan Aral, Lev Muchnik, and Arun Sundararajan.
\newblock {Distinguishing influence-based contagion from homophily-driven
  diffusion in dynamic networks}.
\newblock \emph{Proceedings of the National Academy of Sciences}, 106\penalty0
  (51):\penalty0 21544--21549, 2009.
\newblock \doi{10.1073/pnas.0908800106}.

\bibitem[Arora et~al.(2008)Arora, Telang, and Xu]{arora2008}
Ashish Arora, Rahul Telang, and Hao Xu.
\newblock {Optimal Policy for Software Vulnerability Disclosure}.
\newblock \emph{Management Science}, 54\penalty0 (4):\penalty0 642--656, 2008.
\newblock \doi{10.1287/mnsc.1070.0771}.

\bibitem[Bandura(1977)]{bandura1977}
Albert Bandura.
\newblock \emph{{Social Learning Theory}}.
\newblock Prentice Hall, Englewood Cliffs, NJ, 1977.

\bibitem[Bilge and Dumitra{\c{s}}(2012)]{bilge2012}
Leyla Bilge and Tudor Dumitra{\c{s}}.
\newblock {Before we knew it: an empirical study of zero-day attacks in the
  real world}.
\newblock In \emph{Proceedings of the 2012 ACM conference on Computer and
  communications security}, pages 833--844, 2012.
\newblock \doi{10.1145/2382196.2382284}.

\bibitem[Boyd and Richerson(1985)]{boyd1985}
Robert Boyd and Peter~J. Richerson.
\newblock \emph{{Culture and the Evolutionary Process}}.
\newblock University of Chicago Press, Chicago, 1985.

\bibitem[Broder(1997)]{broder1997}
A.Z. Broder.
\newblock {On the resemblance and containment of documents}.
\newblock In \emph{Proceedings of Compression and Complexity of SEQUENCES
  1997}, pages 21--29, 1997.
\newblock \doi{10.1109/sequen.1997.666900}.

\bibitem[Cameron et~al.(2011)Cameron, Gelbach, and Miller]{cameron2011}
A.~Colin Cameron, Jonah~B. Gelbach, and Douglas~L. Miller.
\newblock {Robust Inference With Multiway Clustering}.
\newblock \emph{Journal of Business \& Economic Statistics}, 29\penalty0
  (2):\penalty0 238--249, 2011.
\newblock \doi{10.1198/jbes.2010.07136}.

\bibitem[Cohen and Levinthal(1990)]{cohen1990}
Wesley~M. Cohen and Daniel~A. Levinthal.
\newblock {Absorptive Capacity: A New Perspective on Learning and Innovation}.
\newblock \emph{Administrative Science Quarterly}, 35\penalty0 (1):\penalty0
  128, 1990.
\newblock \doi{10.2307/2393553}.

\bibitem[Dillon and Tinsley(2008)]{dillon2008}
Robin~L. Dillon and Catherine~H. Tinsley.
\newblock {How Near-Misses Influence Decision Making Under Risk: A Missed
  Opportunity for Learning}.
\newblock \emph{Management Science}, 54\penalty0 (8):\penalty0 1425--1440,
  2008.
\newblock \doi{10.1287/mnsc.1080.0869}.

\bibitem[Ellis and Davidi(2005)]{ellis2005}
Shmuel Ellis and Inbar Davidi.
\newblock {After-Event Reviews: Drawing Lessons From Successful and Failed
  Experience.}
\newblock \emph{Journal of Applied Psychology}, 90\penalty0 (5):\penalty0
  857--871, 2005.
\newblock \doi{10.1037/0021-9010.90.5.857}.

\bibitem[Fehr and G{\"a}chter(2000)]{fehr2000}
Ernst Fehr and Simon G{\"a}chter.
\newblock {Fairness and Retaliation: The Economics of Reciprocity}.
\newblock \emph{Journal of Economic Perspectives}, 14\penalty0 (3):\penalty0
  159--182, 2000.
\newblock \doi{10.1257/jep.14.3.159}.

\bibitem[G{\'a}mez-Guadix et~al.(2015)G{\'a}mez-Guadix, Gini, and
  Calvete]{gamez2015}
Manuel G{\'a}mez-Guadix, Gianluca Gini, and Esther Calvete.
\newblock {Stability of cyberbullying victimization among adolescents:
  Prevalence and association with bully--victim status and psychosocial
  adjustment}.
\newblock \emph{Computers in Human Behavior}, 53:\penalty0 140--148, 2015.
\newblock \doi{10.1016/j.chb.2015.07.007}.

\bibitem[Ganguli et~al.(2022)Ganguli, Lovitt, Kernion, Askell, Bai, Kadavath,
  Mann, Perez, Schiefer, Ndousse, et~al.]{ganguli2022}
Deep Ganguli, Liane Lovitt, Jackson Kernion, Amanda Askell, Yuntao Bai, Saurav
  Kadavath, Ben Mann, Ethan Perez, Nicholas Schiefer, Kamal Ndousse, et~al.
\newblock {Red Teaming Language Models to Reduce Harms: Methods, Scaling
  Behaviors, and Lessons Learned}.
\newblock arXiv preprint arXiv:2209.07858, 2022.

\bibitem[Greshake et~al.(2023)Greshake, Abdelnabi, Mishra, Endres, Holz, and
  Fritz]{greshake2023}
Kai Greshake, Sahar Abdelnabi, Shailesh Mishra, Christoph Endres, Thorsten
  Holz, and Mario Fritz.
\newblock {Not What You've Signed Up For: Compromising Real-World
  LLM-Integrated Applications with Indirect Prompt Injection}.
\newblock In \emph{Proceedings of the 16th ACM Workshop on Artificial
  Intelligence and Security}, pages 79--90, 2023.
\newblock \doi{10.1145/3605764.3623985}.

\bibitem[Holt et~al.(2012)Holt, Bossler, and May]{holt2012}
Thomas~J. Holt, Adam~M. Bossler, and David~C. May.
\newblock {Low Self-Control, Deviant Peer Associations, and Juvenile
  Cyberdeviance}.
\newblock \emph{American Journal of Criminal Justice}, 37\penalty0
  (3):\penalty0 378--395, 2012.
\newblock \doi{10.1007/s12103-011-9117-3}.

\bibitem[Inie et~al.(2025)Inie, Stray, and Derczynski]{inie2025}
Nanna Inie, Jonathan Stray, and Leon Derczynski.
\newblock {Summon a demon and bind it: A grounded theory of LLM red teaming}.
\newblock \emph{PLOS ONE}, 20\penalty0 (1):\penalty0 e0314658, 2025.
\newblock \doi{10.1371/journal.pone.0314658}.

\bibitem[Jennings et~al.(2012)Jennings, Piquero, and Reingle]{jennings2012}
Wesley~G. Jennings, Alex~R. Piquero, and Jennifer~M. Reingle.
\newblock {On the overlap between victimization and offending: A review of the
  literature}.
\newblock \emph{Aggression and Violent Behavior}, 17\penalty0 (1):\penalty0
  16--26, 2012.
\newblock \doi{10.1016/j.avb.2011.09.003}.

\bibitem[Kendal et~al.(2018)Kendal, Boogert, Rendell, Laland, Webster, and
  Jones]{kendal2018}
Rachel~L. Kendal, Neeltje~J. Boogert, Luke Rendell, Kevin~N. Laland, Mike
  Webster, and Patricia~L. Jones.
\newblock {Social Learning Strategies: Bridge-Building between Fields}.
\newblock \emph{Trends in Cognitive Sciences}, 22\penalty0 (7):\penalty0
  651--665, 2018.
\newblock \doi{10.1016/j.tics.2018.04.003}.

\bibitem[Kim and Miner(2007)]{kim2007}
Ji-Yub~(Jay) Kim and Anne~S. Miner.
\newblock {Vicarious Learning from the Failures and Near-Failures of Others:
  Evidence from the U.S. Commercial Banking Industry}.
\newblock \emph{Academy of Management Journal}, 50\penalty0 (3):\penalty0
  687--714, 2007.
\newblock \doi{10.5465/amj.2007.25529755}.

\bibitem[Laland(2004)]{laland2004}
Kevin~N. Laland.
\newblock {Social learning strategies}.
\newblock \emph{Learning \& Behavior}, 32\penalty0 (1):\penalty0 4--14, 2004.
\newblock \doi{10.3758/bf03196002}.

\bibitem[Leukfeldt and Bekkers(2026)]{leukfeldt2026}
E.R. Leukfeldt and L.M.J. Bekkers.
\newblock {From curiosity to crime: Exploring pathways into criminal hacking}.
\newblock \emph{Computers in Human Behavior}, 181:\penalty0 108996, 2026.
\newblock \doi{10.1016/j.chb.2026.108996}.

\bibitem[Lowry et~al.(2017)Lowry, Zhang, and Wu]{lowry2017}
Paul~Benjamin Lowry, Jun Zhang, and Tailai Wu.
\newblock {Nature or nurture? A meta-analysis of the factors that maximize the
  prediction of digital piracy by using social cognitive theory as a
  framework}.
\newblock \emph{Computers in Human Behavior}, 68:\penalty0 104--120, 2017.
\newblock \doi{10.1016/j.chb.2016.11.015}.

\bibitem[Madsen and Desai(2010)]{madsen2010}
Peter~M. Madsen and Vinit Desai.
\newblock {Failing to Learn? The Effects of Failure and Success on
  Organizational Learning in the Global Orbital Launch Vehicle Industry}.
\newblock \emph{Academy of Management Journal}, 53\penalty0 (3):\penalty0
  451--476, 2010.
\newblock \doi{10.5465/amj.2010.51467631}.

\bibitem[Maillart et~al.(2017)Maillart, Zhao, Grossklags, and
  Chuang]{maillart2017}
Thomas Maillart, Mingyi Zhao, Jens Grossklags, and John Chuang.
\newblock {Given enough eyeballs, all bugs are shallow? Revisiting Eric Raymond
  with bug bounty programs}.
\newblock \emph{Journal of Cybersecurity}, 3\penalty0 (2):\penalty0 81--90,
  2017.
\newblock \doi{10.1093/cybsec/tyx008}.

\bibitem[McGregor(2021)]{mcgregor2021}
Sean McGregor.
\newblock {Preventing Repeated Real World AI Failures by Cataloging Incidents:
  The AI Incident Database}.
\newblock \emph{Proceedings of the AAAI Conference on Artificial Intelligence},
  35\penalty0 (17):\penalty0 15458--15463, 2021.
\newblock \doi{10.1609/aaai.v35i17.17817}.

\bibitem[Mesoudi(2011)]{mesoudi2011}
Alex Mesoudi.
\newblock {An experimental comparison of human social learning strategies:
  payoff-biased social learning is adaptive but underused}.
\newblock \emph{Evolution and Human Behavior}, 32\penalty0 (5):\penalty0
  334--342, 2011.
\newblock \doi{10.1016/j.evolhumbehav.2010.12.001}.

\bibitem[Mitra and Ransbotham(2015)]{mitra2015}
Sabyasachi Mitra and Sam Ransbotham.
\newblock {Information Disclosure and the Diffusion of Information Security
  Attacks}.
\newblock \emph{Information Systems Research}, 26\penalty0 (3):\penalty0
  565--584, 2015.
\newblock \doi{10.1287/isre.2015.0587}.

\bibitem[Moneva et~al.(2024)Moneva, Ruiter, and Meinsma]{moneva2024}
Asier Moneva, Stijn Ruiter, and Daniël Meinsma.
\newblock {Criminal expertise and hacking efficiency}.
\newblock \emph{Computers in Human Behavior}, 155:\penalty0 108180, 2024.
\newblock \doi{10.1016/j.chb.2024.108180}.

\bibitem[Montgomery et~al.(2018)Montgomery, Nyhan, and Torres]{montgomery2018}
Jacob~M. Montgomery, Brendan Nyhan, and Michelle Torres.
\newblock {How Conditioning on Posttreatment Variables Can Ruin Your Experiment
  and What to Do about It}.
\newblock \emph{American Journal of Political Science}, 62\penalty0
  (3):\penalty0 760--775, 2018.
\newblock \doi{10.1111/ajps.12357}.

\bibitem[Morris and Blackburn(2009)]{morris2009}
Robert~G. Morris and Ashley~G. Blackburn.
\newblock {Cracking the Code: An Empirical Exploration of Social Learning
  Theory and Computer Crime}.
\newblock \emph{Journal of Crime and Justice}, 32\penalty0 (1):\penalty0 1--34,
  2009.
\newblock \doi{10.1080/0735648x.2009.9721260}.

\bibitem[Perez and Ribeiro(2022)]{perez2022}
F{\'a}bio Perez and Ian Ribeiro.
\newblock {Ignore Previous Prompt: Attack Techniques For Language Models}.
\newblock NeurIPS ML Safety Workshop, arXiv:2211.09527, 2022.

\bibitem[Rendell et~al.(2010)Rendell, Boyd, Cownden, Enquist, Eriksson,
  Feldman, Fogarty, Ghirlanda, Lillicrap, and Laland]{rendell2010}
L.~Rendell, R.~Boyd, D.~Cownden, M.~Enquist, K.~Eriksson, M.~W. Feldman,
  L.~Fogarty, S.~Ghirlanda, T.~Lillicrap, and K.~N. Laland.
\newblock {Why Copy Others? Insights from the Social Learning Strategies
  Tournament}.
\newblock \emph{Science}, 328\penalty0 (5975):\penalty0 208--213, 2010.
\newblock \doi{10.1126/science.1184719}.

\bibitem[Schulhoff et~al.(2023)Schulhoff, Pinto, Khan, Bouchard, Si, Anati,
  Tagliabue, Kost, Carnahan, and Boyd-Graber]{schulhoff2023}
Sander Schulhoff, Jeremy Pinto, Anaum Khan, Louis-Fran{\c{c}}ois Bouchard,
  Chenglei Si, Svetlina Anati, Valen Tagliabue, Anson Kost, Christopher
  Carnahan, and Jordan Boyd-Graber.
\newblock {Ignore This Title and HackAPrompt: Exposing Systemic Vulnerabilities
  of LLMs Through a Global Prompt Hacking Competition}.
\newblock In \emph{Proceedings of the 2023 Conference on Empirical Methods in
  Natural Language Processing}, pages 4945--4977, 2023.
\newblock \doi{10.18653/v1/2023.emnlp-main.302}.

\bibitem[Shalizi and Thomas(2011)]{shalizi2011}
Cosma~Rohilla Shalizi and Andrew~C. Thomas.
\newblock {Homophily and Contagion Are Generically Confounded in Observational
  Social Network Studies}.
\newblock \emph{Sociological Methods \& Research}, 40\penalty0 (2):\penalty0
  211--239, 2011.
\newblock \doi{10.1177/0049124111404820}.

\bibitem[Shen et~al.(2024)Shen, Chen, Backes, Shen, and Zhang]{shen2024dan}
Xinyue Shen, Zeyuan Chen, Michael Backes, Yun Shen, and Yang Zhang.
\newblock {``Do Anything Now'': Characterizing and Evaluating In-The-Wild
  Jailbreak Prompts on Large Language Models}.
\newblock In \emph{Proceedings of the 2024 on ACM SIGSAC Conference on Computer
  and Communications Security}, pages 1671--1685, 2024.
\newblock \doi{10.1145/3658644.3670388}.

\bibitem[{The PyFixest Authors}(2026)]{pyfixest}
{The PyFixest Authors}.
\newblock {pyfixest: Fast high-dimensional fixed effect estimation in
  {Python}}.
\newblock Python package, version 0.60.0, 2026.
\newblock URL \url{https://github.com/py-econometrics/pyfixest}.

\bibitem[Toyer et~al.(2024{\natexlab{a}})Toyer, Watkins, Mendes, Svegliato,
  Bailey, Wang, Ong, Elmaaroufi, Abbeel, Darrell, Ritter, and
  Russell]{toyer2024}
Sam Toyer, Olivia Watkins, Ethan~Adrian Mendes, Justin Svegliato, Luke Bailey,
  Tiffany Wang, Isaac Ong, Karim Elmaaroufi, Pieter Abbeel, Trevor Darrell,
  Alan Ritter, and Stuart Russell.
\newblock {Tensor Trust: Interpretable Prompt Injection Attacks from an Online
  Game}.
\newblock In \emph{The Twelfth International Conference on Learning
  Representations (ICLR)}, pages 18714--18746, 2024{\natexlab{a}}.

\bibitem[Toyer et~al.(2024{\natexlab{b}})]{tensortrustdata}
Sam Toyer et~al.
\newblock {The {Tensor Trust} dataset (benchmarks v1, raw data dump v2)}.
\newblock Data repository, 2024{\natexlab{b}}.
\newblock URL \url{https://github.com/HumanCompatibleAI/tensor-trust-data}.

\bibitem[Vassilev et~al.(2025)Vassilev, Oprea, Fordyce, Anderson, Davies, and
  Hamin]{vassilev2025}
Apostol Vassilev, Alina Oprea, Alie Fordyce, Hyrum Anderson, Xander Davies, and
  Maia Hamin.
\newblock {Adversarial Machine Learning: A Taxonomy and Terminology of Attacks
  and Mitigations}.
\newblock Technical Report NIST AI 100-2e2025, National Institute of Standards
  and Technology, 2025.
\newblock \doi{10.6028/nist.ai.100-2e2025}.

\bibitem[Virgara and Whitten(2023)]{virgara2023}
Josephine~L. Virgara and Tyson Whitten.
\newblock {A systematic literature review of the longitudinal risk factors
  associated with juvenile cyber-deviance}.
\newblock \emph{Computers in Human Behavior}, 141:\penalty0 107613, 2023.
\newblock \doi{10.1016/j.chb.2022.107613}.

\bibitem[Wei et~al.(2023)Wei, Haghtalab, and Steinhardt]{wei2023}
Alexander Wei, Nika Haghtalab, and Jacob Steinhardt.
\newblock {Jailbroken: How Does {LLM} Safety Training Fail?}
\newblock In \emph{Advances in Neural Information Processing Systems 36
  (NeurIPS)}, pages 80079--80110, 2023.

\bibitem[Weulen~Kranenbarg et~al.(2019)Weulen~Kranenbarg, Holt, and van
  Gelder]{kranenbarg2019}
Marleen Weulen~Kranenbarg, Thomas~J. Holt, and Jean-Louis van Gelder.
\newblock {Offending and Victimization in the Digital Age: Comparing Correlates
  of Cybercrime and Traditional Offending-Only, Victimization-Only and the
  Victimization-Offending Overlap}.
\newblock \emph{Deviant Behavior}, 40\penalty0 (1):\penalty0 40--55, 2019.
\newblock \doi{10.1080/01639625.2017.1411030}.

\bibitem[Yokotani and Takano(2021)]{yokotani2021}
Kenji Yokotani and Masanori Takano.
\newblock {Social contagion of cyberbullying via online perpetrator and victim
  networks}.
\newblock \emph{Computers in Human Behavior}, 119:\penalty0 106719, 2021.
\newblock \doi{10.1016/j.chb.2021.106719}.

\bibitem[Yu et~al.(2024)Yu, Liu, Liang, Cameron, Xiao, and Zhang]{yu2024}
Zhiyuan Yu, Xiaogeng Liu, Shunning Liang, Zach Cameron, Chaowei Xiao, and Ning
  Zhang.
\newblock {Don't Listen To Me: Understanding and Exploring Jailbreak Prompts of
  Large Language Models}.
\newblock In \emph{Proceedings of the 33rd USENIX Security Symposium}, pages
  4675--4692, 2024.

\end{thebibliography}

\end{document}